\documentclass[amsmath,amssymb,twocolumn,pra]{revtex4-1}
\usepackage{bbm}
\usepackage{amssymb}

\usepackage{epsfig}
\usepackage{graphicx}
\usepackage{mathrsfs}
\usepackage{amsfonts}
\usepackage[english]{babel}
\usepackage{color}

\begin{document}

\title{ Unveiling quantum entanglement and correlation of sub-Ohmic and Ohmic baths for quantum phase transitions in dissipative systems}

\author{Xiaohui Qian}
\author{Nengji Zhou}
\email[Corresponding author:~]{zhounengji@hznu.edu.cn}
\author{Sun Zhe}
\email[Corresponding author:~]{sunzhe@hznu.edu.cn}
\date{\today}
\affiliation{School of Physics, Hangzhou Normal University, Hangzhou 311121, China
}

\begin{abstract}
By employing the spin-boson model in a dense limit of environmental modes, quantum entanglement and correlation of sub-Ohmic and Ohmic baths for dissipative quantum phase transitions are numerically investigated based on the variational principle. With several measures borrowed from quantum information theory, three different types of singularities are found for the first-order, second-order, and Kosterlitz-Thouless phase transitions, respectively, and the values of transition points and critical exponents are accurately determined. Besides, the scaling form of the quantum discord in the Ohmic case is identified, quite different from that in the sub-Ohmic regime. In two-spin model, two distinct behaviors of the quantum discord are uncovered: one is related to the quantum entanglement between two spins, and the other is decided by the correlation function in the position space rather than the entanglement.

\end{abstract}

\maketitle

\section{Introduction}
As a new developed interdisciplinary field in recent decades, quantum information science mainly based on the principles of quantum mechanics has attracted much attention and developed quickly in both theoretical and experimental researches \cite{bou00,nie00,bra05}.  At the core of the quantum information, quantum entanglement \cite{ami08,hor09} which is an important resource in the processes of realizing quantum computation and quantum information tasks, has been extensively studied, leading to powerful applications. Examples are coupled electronic and vibrational motions in molecules \cite{ish10,mck15}, quantum teleportation \cite{ben93}, quantum key distribution \cite{jen00}, quantum dense coding \cite{ben92}, telecloning \cite{mur99}, one-way quantum computing \cite{rau01}, and quantum estimation \cite{ven07}. However, it is very fragile and easily destroyed by the decoherence effect from the surrounding environment \cite{hor09,ber17}. Hence, a full understanding of the influences of environmental noise on the quantum entanglement is very important, though it is a long standing challenge in the study of open quantum systems.

A wide variety of characterizations for quantumness have been proposed for the investigation of quantum phase transitions describing sudden changes of the many-body ground state as a nonthermal control parameter moves through the critical value at zero temperature \cite{vid03,qua06, wer10,che16,hu20,tan20,wu21}. These quantum information oriented methods have a common advantage in that they can study phase transitions without any knowledge of order parameters and symmetries in priority. Quantum entanglement is one of the most famous indicators \cite{ost02}, and quantum discord which reflects the quantum correlation between two components of the system, complements the entanglement in certain situations to detect phase transitions \cite{ber17}.

Spin-boson model (SBM) is a well-known dissipative model describing the interaction between a qubit (spin or two-level system) and an infinite collection of harmonic oscillators (bosonic bath) \cite{leg87,hur08,bre16}. In spite of apparent simplicity, it has been widely used to study dephasing and dissipative effects from the environment \cite{wal16,wu17,hur18,str18,mas20}. Recent studies show that there exists a localized-delocalized quantum phase transition, as long as the coupling between the system and environment is characterized by a continuous spectral function $J(\omega) \propto  \omega^s$ \cite{voj05,hur08}. A rich phase diagram has been found for different values of the spectral exponent $s$ \cite{chi11,win14,ber14,guo12,zho14,wan19b}. More specifically, the phase transition is of second order in the sub-Ohmic regime ($s < 1$), and of the Kosterlitz-Thouless type in the Ohmic case($s =1$). In the super-Ohmic regime, however, there is no phase transition. Very recently, the transition has been inferred to be of first order even in the Ohmic case, for the two-spin SBM with a strong antiferromagnetic spin-spin coupling \cite{zho18}. It is in contrast to that in the absence of spin-spin coupling where the transition still belongs to the Kosterlitz-Thouless universality class. However, previous studies principally focused on the spin-related observations, such as the spontaneous spin magnetization and the spin coherence as well as the von Neumann entropy characterizing the system-environment entanglement \cite{hur07,naz12,guo12,fre13,bru14,zho14,wan20}. Bath-related observables which can provide a direct measurement of the quantum criticality intrinsic to the environment were less considered \cite{blu17,zho18,zho21}.

Besides, SBM quantum simulation schemes have been realized in recent experiments of superconducting quantum circuits \cite{lep18,mag18,yam19} and trapped atomic ion crystals \cite{por08,lem18}. It will become feasible in the near future to experimentally measure quantum entanglement and correlation of sub-Ohmic and Ohmic baths. As a result, it crucially requires a deepened knowledge of the critical behaviors nearby the phase transition, but up to now the progress is limited.  Although much effort has been devoted to the study of the quantum entanglement in discrete variable systems \cite{ost02,lar05,cho06,ami08}, whether the scaling law still holds for continuous variables remains an open issue. Being one of the simplest prototypes of continuous variable bipartite systems, two-mode Gaussian state defined as the set of states with Gaussian characteristic functions and quasi-probability distributions has aroused great interest \cite{ade10,tse17,isa17}. Using covariance matrices and symplectic analysis, numerical investigations are carried out on the quantum entanglement and correlation within the bosonic bath.

This paper aims at a comprehensive study of quantum entanglement and correlation of sub-Ohmic and Ohmic baths for different types of quantum phase transitions, such as first-order, second-order and Kosterlitz-Thouless transitions, taking the single-spin and two-spin SBM as examples. To obtain an accurate description of the ground state for both the spin system and its environment, numerical variational method (NVM)  \cite{zho14,blu17} based on systematic coherent-state decomposition is employed here, which has been proved to be valid in tackling ground-state phase transitions and quantum dynamics \cite{zho15,zho15b,zho16,wan16, wan17}. Moreover, the quantum entanglement and correlation between two spins are also measured in the two-spin model, and critical behaviors of them are carefully analyzed, in comparison with those within the bosonic bath. The rest of the paper is organized as follows. In section~\ref{sec:mod}, the models and variational approach are described. In section~\ref{sec:num}, numerical results of quantum entanglement and correlation as well as the derivative of the ground-state energy are presented for quantum phase transitions in dissipative systems involving the single-spin and two-spin SBM in the sub-Ohmic/Ohmic regime. Finally, conclusions are drawn in section~\ref{sec:con}.

\section{Model and Method}\label{sec:mod}
The Hamiltonians of the single-spin and two-spin SBM are given by
\begin{equation}
\label{Ohami_single}
\hat{H} =  \frac{\varepsilon}{2}\sigma_z-\frac{\Delta}{2}\sigma_x + \sum_{k} \omega_k b_{k}^\dag b_{k}  +  \frac{\sigma_z}{2}\sum_k \lambda_k(b^\dag_{k}+b_{k}),
\end{equation}
and
\begin{eqnarray}
\label{Ohami_two}
\hat{H} & = & \frac{\varepsilon}{2}(\sigma_{z1}+\sigma_{z2})-\frac{\Delta}{2}(\sigma_{x1}+\sigma_{x2}) + \sum_{k} \omega_k b_{k}^\dag b_{k} \nonumber \\
& + & \frac{\sigma_{z1}+\sigma_{z2}}{2}\sum_k \lambda_k(b^\dag_{k}+b_{k}) + \frac{K}{4}\sigma_{z1}\sigma_{z2},
\end{eqnarray}
respectively, where $\varepsilon$ ($\Delta$) denotes the energy bias (bare tunneling amplitude), $b^\dag_k$ ($b_k$) represents the bosonic creation (annihilation) operator of the $k$-th bath mode with the frequency $\omega_k$, $\sigma_x$ and $\sigma_z$ are Pauli spin-$1/2$ operators, and $\lambda_k$ signifies the coupling amplitude between the system and bath. The subscripts of $\sigma_{xi}$ and $\sigma_{zi}~(i=1, 2)$ in Eq.~(\ref{Ohami_two}) correspond to the $1$th and $2$th spins, respectively, and $K$ represents the Ising-type spin-spin interaction. Here we state the notation $\varepsilon$ only for completeness, and the focus of the paper lies on the case $\varepsilon=0$.

The parameters $\lambda_k$ and  $\omega_k$ are obtained from the spectral density function $J(\omega)=2\alpha\omega_c^{1-s}\omega^s =\sum_k\lambda_k^2\delta(\omega-\omega_k)$ \cite{bul05, voj05,zha10,zho14, blu17},
\begin{equation}
\label{sbm1_dis}
\lambda_k^2  =  \int^{\Lambda_{k+1}\omega_c}_{\Lambda_k\omega_c}dt J(t), \quad \omega_k  =  \lambda^{-2}_k \int^{\Lambda_{k+1}\omega_c}_{\Lambda_k\omega_c}dtJ(t)t,
\end{equation}
where $\alpha$ denotes the dimensionless coupling strength, $\omega_c$ represents the high-frequency cutoff, and $\Lambda_k=\Lambda^{k-M}$ is set with the factor $\Lambda$ in the logarithmic discretization procedure \cite{bul03,hur08,chi11,fre13}. In this work, the continuum limit $\Lambda \rightarrow 1$ is required in order to obtain an accurate description of the ground state for the SBM with a high dense spectrum. To simplify notations, hereafter we fix the maximum frequency $\omega_c=1$. Other model parameters which are in unit of $\omega_c$, i.e., $\varepsilon, \Delta$, and $K$, are then set to be dimensionless.

The trial ansatz composed of a systematic coherent-state expansion, termed as the ``Davydov multi-$D_1$ ansatz'' \cite{zho14,zho15,wan16}, is used in variational calculations,
\begin{eqnarray}
\label{vmwave1}
|\Psi_1 \rangle & = & | + \rangle \sum_{n=1}^{N} A_n \exp\left[ \sum_{k=1}^{M}\left(f_{n,k}b_k^{\dag} - \mbox{H}.\mbox{c}.\right)\right] |0\rangle_{\textrm{b}} \nonumber \\
              & + & |- \rangle \sum_{n=1}^{N} D_n \exp\left[ \sum_{k=1}^{M}\left(g_{n,k}b_k^{\dag} - \mbox{H}.\mbox{c}.\right)\right] |0\rangle_{\textrm{b}},
\end{eqnarray}
for the single-spin SBM, and
\begin{eqnarray}
\label{vmwave2}
|\Psi_2 \rangle & = & |++\rangle \sum_{n=1}^{N} A_n \exp\left[ \sum_{k=1}^{M}\left(f_{n,k}b_k^{\dag} - \mbox{H}.\mbox{c}.\right)\right] |0\rangle_{\textrm{b}} \nonumber \\
& + & |+-\rangle \sum_{n=1}^{N} B_n \exp\left[ \sum_{k=1}^{M}\left(g_{n,k}b_k^{\dag} - \mbox{H}.\mbox{c}.\right)\right]
|0\rangle_{\textrm{b}}   \nonumber\\
& + & |-+\rangle \sum_{n=1}^{N} C_n \exp\left[ \sum_{k=1}^{M}\left(h_{n,k}b_k^{\dag} - \mbox{H}.\mbox{c}.\right)\right]
|0\rangle_{\textrm{b}} \\
& + & |-- \rangle \sum_{n=1}^{N} D_n \exp\left[ \sum_{k=1}^{M}\left(p_{n,k}b_k^{\dag} - \mbox{H}.\mbox{c}.\right)\right]
|0\rangle_{\textrm{b}},\nonumber
\end{eqnarray}
for the two-spin SBM, respectively. In Eqs.~(\ref{vmwave1}) and (\ref{vmwave2}), H$.$c$.\!$ denotes Hermitian conjugate, $+$ ($-$) stands for the spin up (down) state, and $|0\rangle_{\rm b}$ is the vacuum state of the bosonic bath. The variational parameters $f_{n,k},~ g_{n,k},~ h_{n,k}$, and $p_{n,k}$ represent the displacements of the coherent states correlated to the spin configurations, and $A_n,~ B_n,~ C_n$, and $D_n$ are weights of the coherent states. The subscripts $n$ and $k$ correspond to the ranks of the coherent superposition state and effective bath mode, respectively. In fact, these trial wavefunctions above are generalized Silbey-Harris Ansatz based on the work of Luther and Emery in the $1970$s, and of Silbey and Harris at $1984$ \cite{eme74, sil84}.

The ground state  $|\Psi_g\rangle$ is determined by minimizing the energy expressed as $E=\mathcal{H}/\mathcal{N}$ using the Hamiltonian expectation $\mathcal{H}=\langle \Psi_{1,2}|\hat{H}|\Psi_{1,2}\rangle$ and the norm of the wave function $\mathcal{N}=\langle \Psi_{1,2} |\Psi_{1,2}\rangle$. The variational procedure entails a set of self-consistency equations which can be numerical solved with the relaxation iteration technique and global optimization algorithm. For each set of the model parameters $(\alpha, \Delta, \Lambda, K)$, more than one hundred random initial states are taken to reduce statistical noise. Since time evolutions of variational parameters in the iteration are dependent on the initial states, the relaxation dynamics is considered to be not universal.

With the ground-state wavefunction at hand, two phase-space variables are given by $x_k=\langle\Psi_g|\hat{x}_k|\Psi_g\rangle$ and $p_k=\langle\Psi_g|\hat{p}_k|\Psi_g\rangle$ as the expectation values of the position and momentum for the k-th bath mode, respectively, where $\hat{x}_k$ and $\hat{p}_k$ are
\begin{equation}
\label{vm_xp}
\hat{x}_k = \left(b_k+b_k^{\dag}\right)/\sqrt{2}, \qquad \hat{p}_k = i\left(b_k^{\dag}-b_k\right)/\sqrt{2}.
\end{equation}
The variances of phase-space variables and correlation functions are then measured,
\begin{eqnarray}
\label{phase var}
\Delta X_{\rm b} & = & \langle \Psi_{\rm g}|(\hat{x}_k)^2 |\Psi_{\rm g}\rangle - \langle \Psi_{\rm g}|\hat{x}_k|\Psi_{\rm g}\rangle^2, \nonumber \\
\Delta P_{\rm b} &= &\langle \Psi_{\rm g}|(\hat{p}_k)^2 |\Psi_{\rm g}\rangle,  \nonumber \\
{\rm Cor_X} & = & \langle \Psi_{\rm g}|\hat{x}_k \hat{x}_l |\Psi_{\rm g}\rangle -  \langle \Psi_{\rm g}|\hat{x}_k|\Psi_{\rm g}\rangle  \langle \Psi_{\rm g}|\hat{x}_l|\Psi_{\rm g}\rangle, \nonumber \\
{\rm Cor_P} & = & \langle \Psi_{\rm g}|\hat{p}_k \hat{p}_l |\Psi_{\rm g}\rangle,
\end{eqnarray}
where the subscripts $k$ and $l$ correspond to the $k$-th and \break $l$-th bath modes, respectively, and $\langle \Psi_{\rm g}|\hat{p}_k|\Psi_{\rm g}\rangle \equiv 0$ is expected for the ground state.

Since the number of coherent-superposition $N$ in Eqs.~(\ref{vmwave1}) and (\ref{vmwave2}) is small, the ground-state state of each bath mode can be approximated to a Gaussian state. By that means, it can be characterized by first and second statistical moments, denoted by the vector $\overrightarrow{r}=(x_k,p_k,x_l,p_l)$ and its covariance matrix $\sigma_{ij}=\langle \Psi_{\rm g}|\hat{r}_i\hat{r}_j|\Psi_{\rm g}\rangle - \langle\Psi_{\rm g} \hat{r}_i|\Psi_{\rm g}\rangle \langle\Psi_g| \hat{r}_j|\Psi_g\rangle$, respectively. In terms of the variances of phase-space variables, the latter can also be written as
\begin{equation}
\left(
  \begin{array}{cccc}
    \Delta X_k & 0 & {\rm Cor_X} & 0 \\
    0          & \Delta P_k & 0 &  {\rm Cor_P} \\
   {\rm Cor_X} & 0 & \Delta X_l & 0 \\
    0          & \Delta {\rm Cor_P} & 0 &  \Delta P_l  \\
  \end{array}
\right) =
\left(
  \begin{array}{cc}
    A & C \\
    C^{T} & B \\
    \end{array}
\right),
\label{cov_matrix}
\end{equation}
where $A,B$ and $C$ are three $2\times2$ matrices, and $C^{T}$ represents the transpose of the matrix $C$.

For an arbitrary two-mode Gaussian state $\sigma$, the von-Neumann entropy $S_{\rm b}$ is measured firstly,
\begin{equation}
\label{entropy}
S_{\rm b}=f\left(n_{-}\right)+f\left(n_{+}\right),
\end{equation}
where $n_{\pm}$ is given by
\begin{equation}
\label{n_plus}
n_{\pm}^{2}=\frac{\Delta\pm\sqrt{\Delta^{2}-4{\rm det}\sigma}}{2},
\end{equation}
with the invariant $\Delta={\rm det}A+{\rm det}B+2{\rm det}C$ under the action of the symplectic transformation, and the form of the function $f(x)$ is given by
\begin{equation}
f\left(x\right)=\left(x+\frac{1}{2}\right)\ln\left(x+\frac{1}{2}\right)-\left(x-\frac{1}{2}\right)\ln\left(x-\frac{1}{2}\right).
\end{equation}
The determinant of the covariance matrix $\sigma$ and its submatrix $A,B$, and $C$ are calculated as
\begin{eqnarray}
{\rm det}A	&=&	\Delta X_{k}\Delta P_{k},  \nonumber \\
{\rm det}B	&=&	\Delta X_{l}\Delta P_{l},  \nonumber \\
{\rm det}C	&=&	\rm {\rm Cor}_{x}{\rm Cor}_{p},   \\
{\rm det}\sigma&=&\Delta X_{k}\Delta P_{k}\Delta X_{l}\Delta P_{l}+{\rm Cor}_x{\rm Cor}_x{\rm Cor}_p{\rm Cor}_p \nonumber \\
&-&\Delta X_{k}\Delta X_{l}{\rm Cor}_p{\rm Cor}_p-\Delta P_{k}\Delta P_{l}{\rm Cor}_x{\rm Cor}_x. \nonumber
\end{eqnarray}
Another measurement referred to as the linear entropy, $S_{\rm L,b}$, is also carried out,
\begin{equation}
S_{\rm L,b} =  1 - \mu_{\rm b}= 1 - {\rm Tr}\left[\rho^{2}\right]=1 - 1/\left(4\sqrt{{\rm det}\sigma}\right),
\label{purity}
\end{equation}
where $\mu_{\rm b}$ denotes the purity ranging from $1$ for pure states to the limiting value $0$ for completely mixed states, since no finite lower bound to the $2$-norm of $\rho$ exists due to the infinite dimension of the Hilbert space \cite{ade06}. Besides, the mutual information is also investigated,
\begin{eqnarray}
I_{\rm b} &= & S_{\rm b}\left(\sigma_{1}\right)+S_{\rm b}\left(\sigma_{2}\right)-S_{\rm b}\left(\sigma\right) \nonumber \\
&=&f\left(a\right)+f\left(b\right)-f\left(n_{-}\right)-f\left(n_{+}\right),
\end{eqnarray}
where $\sigma_{1,2}$ denotes the reduced state of the subsystem $1$ or $2$, and the parameters $a$ and $b$ represent $\sqrt{{\rm det}A}$ and $\sqrt{{\rm det}B}$, respectively.

In order to quantify the degree of entanglement, the logarithmic negativity $E_{\rm N,b}={\rm max}\{0,-\log_{2}2\tilde{\nu}_{-}\}$ is introduced , where $\tilde{\nu}_{-}$ is the symplectic eigenvalues of the partial transpose $\tilde{\sigma}$ of the two-mode covariance matrix
\begin{equation}
\tilde{\nu}_{\pm}^{2}	=	\frac{\tilde{\Delta}\pm\sqrt{\tilde{\Delta}^{2}-4{\rm det}\sigma}}{2},
\end{equation}
where $\tilde{\Delta}={\rm detA+{\rm det}B-2{\rm det}C}$ is another symplectic invariant.

Finally, quantum discord $D_{\rm b}=I_{\rm b}-C_{\rm b}$ is considered as a measure of all nonclassical correlations in a bipartite state, including but not restricted to entanglement. For pure entangled states, quantum discord coincides with the entropy of entanglement. It also can be nonvanishing for some mixed separable state wherein the correlation depicted by the positive discord is an indicator of quantumness. States with zero discord represent essentially a classical probability distribution embedded in a quantum system. The classical correlation $C_{\rm b}$ is measured by maximizing over all possible measurements, taking the form
\begin{eqnarray}
\label{class_correlation}
C_{\rm b}\left(\sigma\right)	&=&	S_{\rm b}\left(\sigma_{1}\right)-{\rm inf}_{\{\prod_{i}\}}\{S_{\rm b}\left(\sigma_{1|2}\right)\} \nonumber \\
	&=&	S_{\rm b}\left(\sigma_{1}\right)-{\rm inf}_{\{\prod_{i}\}}\sum_{i}p_{i}S_{\rm b}\left(\rho_{1i}\right),
\end{eqnarray}
where $p_{i}$ is the measurement probability for the i-th local projector, and $\rho_{1i}$ denotes the reduced state of subsystem $1$ after local measurements.

As reported in previous work \cite{ade10,isa17}, the quantum discord of a general two-mode Gaussian state is given by $D_{\rm b}\left(\sigma\right) =f\left(b\right)+f\left(e\right)-f\left(n_{-}\right)-f\left(n_{+}\right)$, where $b,n_{-},n_{+}$, and $f\left(x\right)$ are already mentioned before, and the value of $e=\sqrt{{\rm det}\epsilon}$ is estimated by
\begin{equation}
\label{discord_bosonic}
{\rm det}\epsilon=\begin{cases}
\frac{2\gamma^{2}+\left(\beta-1\right)\left(\delta-\alpha\right)+2\left|\gamma\right|\sqrt{\gamma^{2}+\left(\beta-1\right)\left(\delta-\alpha\right)}}{4\left(\beta-1\right)^{2}} & {\rm if}\Gamma \geq 0,\\
\\
\frac{\alpha\beta-\gamma^{2}+\delta-\sqrt{\gamma^{4}+\left(\delta-\alpha\beta\right)^{2}-2\gamma^{2}\left(\alpha\beta+\delta\right)}}{8\beta} & {\rm if} \Gamma < 0,
\end{cases}
\end{equation}
with $\Gamma=\left(1+\beta\right)\gamma^{2}\left(\alpha+\delta\right)-\left(\delta-\alpha\beta\right)^{2}, \alpha=4\det A,\beta=4{\rm det}B,\gamma=4{\rm det}C$, and $\delta=16{\rm det}\sigma$.

Since the quantum entanglement and correlation defined in Eqs.(\ref{entropy})-(\ref{discord_bosonic}) can be investigated for arbitrary two bath modes, without losing any generality we fix one frequency $\omega_{l}=\omega_c=1$ in the following. Taking their summations over the other frequency $\omega_k$, efficient indicators $\sum \rm Cor_X, \sum S_{\rm b}, \sum \rm I_{\rm b}, \sum S_{\rm L,b}, \sum E_{\rm N,b}$, and $\sum D_{\rm b}$ are thus obtained for quantum phase transitions.

\section{Numerical results}\label{sec:num}
By means of the entanglement, correlation, and entropy as well as the derivative of the ground-state energy, ground-state phase transitions in single-spin and two-spin SBM are numerically investigated based on the variational principle. In light of rich phase diagrams, four different cases are considered as examples, which are the single-spin one in sub-Ohmic regime, Ohmic ones in single-spin and two-spin models, respectively, and two-spin one with a strong antiferromagnetic coupling. Correspondingly, three different types of singularities and critical behaviors are identified for second-order, Kosterlitz-Thouless, and first-order transitions.

\subsection{Second-order phase transitions}
Firstly, the localized-delocalized phase transition in the sub-Ohmic SBM is demonstrated by setting the model parameters $s=0.2, \Delta=0.1$, and $\varepsilon=0$. Convergence test of variational results is performed against the number of the coherent-superposition states $N$, and it is concluded that $N=8$ is sufficient in variational method to accurately describe ground states.

\subsubsection{ Derivative of ground-state energy}
\begin{figure*}[htb]
\centering
\epsfysize=7cm \epsfclipoff \fboxsep=0pt
\setlength{\unitlength}{1.cm}
\begin{picture}(10,7)(0,0)
\put(-3.0, 0.0){{\epsffile{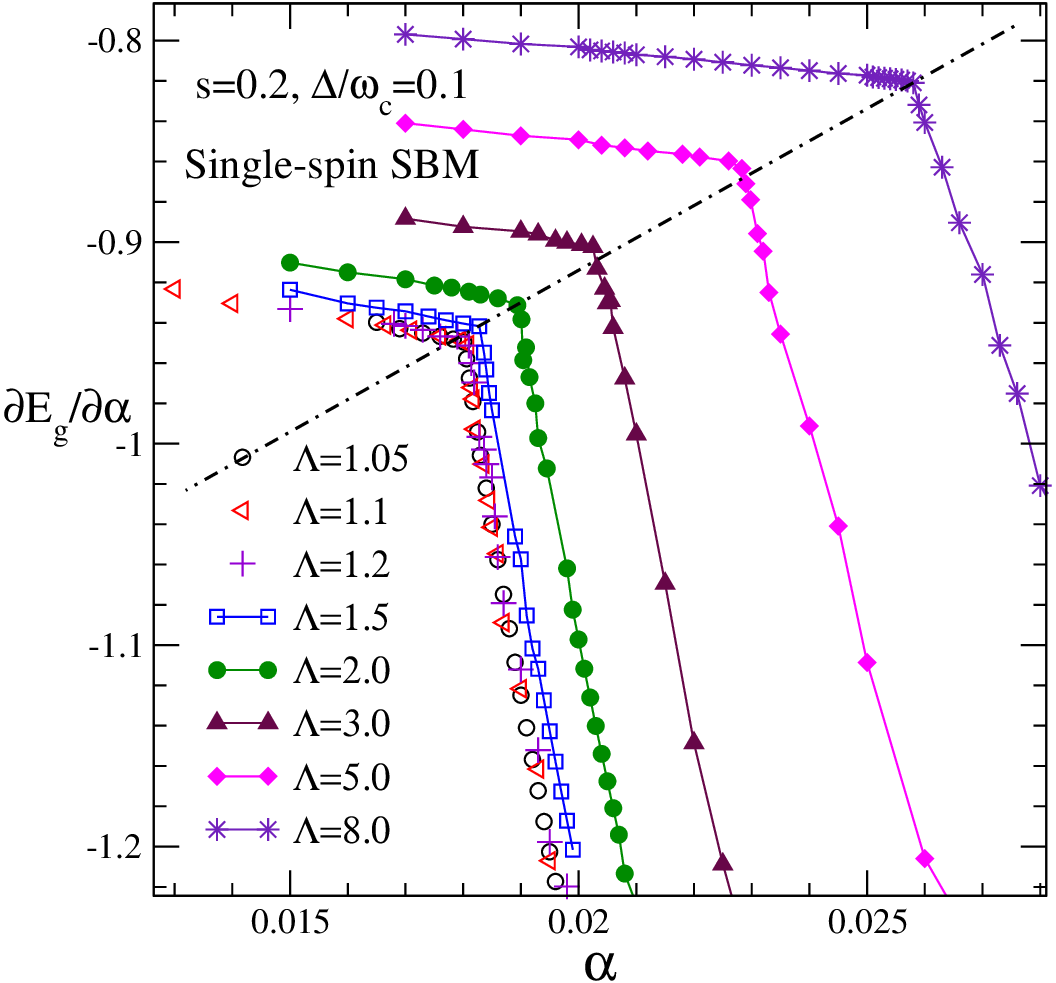}}}\epsfysize=7cm
\put(5.2, 0.0){{\epsffile{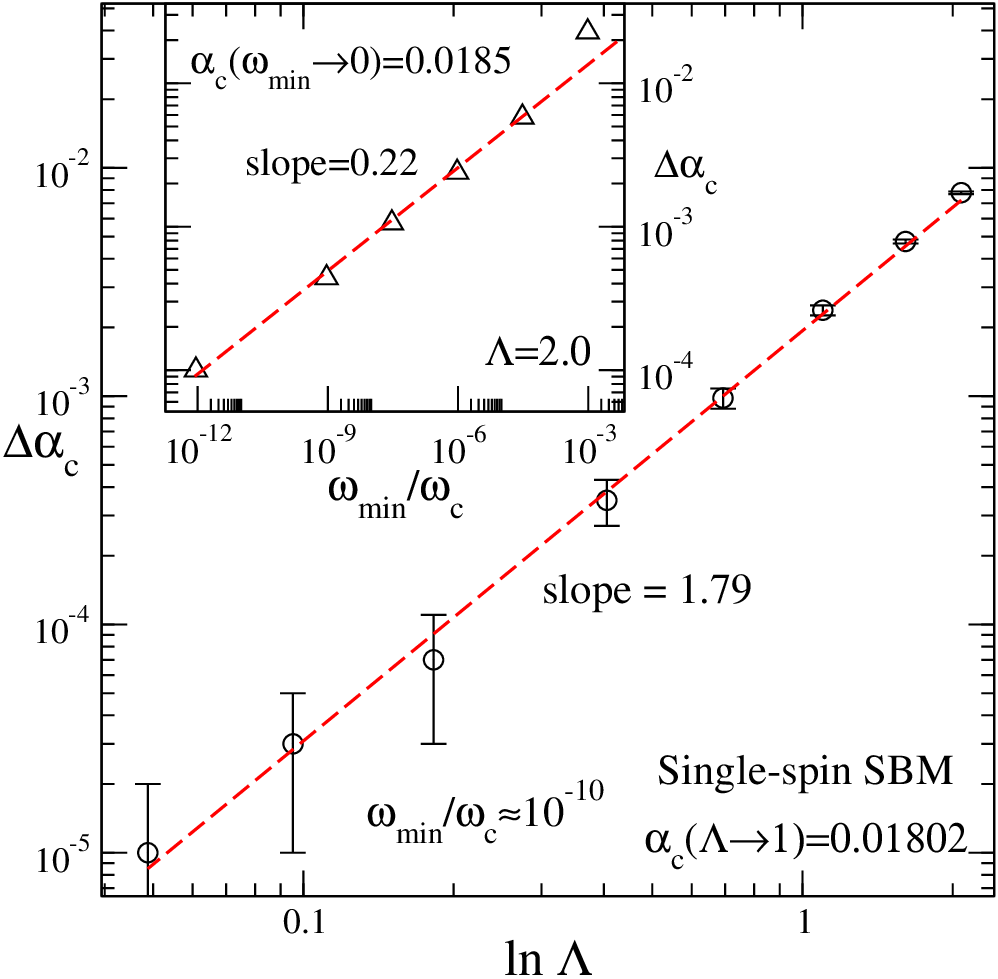}}}
\end{picture}

\hspace{0.0cm}\footnotesize{(a)}\hspace{8.0cm}\footnotesize{(b)}
\caption{(a) The first derivative of the ground-state energy $E_g$ in the sub-Ohimc SBM ($s=0.2$) is displayed as a function of the coupling strength $\alpha$ for various values of the discretization factor $\Lambda$ with the same lowest frequency $\omega_{\rm min}=\Lambda^{-M} \omega_c\approx 10^{-10}\omega_c$. The tunneling constant $\Delta=0.1\omega_c$ and the number of coherent-superposition states $N=8$ are set. The phase boundary where the slope is discontinuous is marked by the dash-dotted line. (b) The shift of the transition point $\Delta\alpha_c$ with respect to $\ln \Lambda$ and $\omega_{\rm min}$ is plotted with open circles and open triangles (in the inset), respectively. Dashed lines represent power-law fits.}
\label{f1}
\end{figure*}

In Fig.~\ref{f1}(a), the first derivative of the ground-state energy, $\partial E_g/\partial \alpha$, is plotted against the coupling $\alpha$. To investigate the discretization effect, one employs different values of the discretization factors $\Lambda=1.05 \sim 8$ with the same lowest frequency $\omega_{\rm min}\approx 10^{-10}$. All curves  decrease with the coupling $\alpha$, and have sharp kinks, indicating a high-order singularity expected at the transition point. It confirms that the phase transition is of second order. Moreover, the transition boundary marked by the dash-dotted line displays a linear dependence of  $\partial E_g/\partial \alpha$ on the critical coupling $\alpha_c$. The convergence is reached at $\Lambda=1.05$, corresponding to the number of effective bath modes $M=430$. It is about the same order of magnitude as that in the Davydov work on the quantum dynamics of the SBM \cite{har19}.

Taking the asymptotic value $\alpha_{c,\Lambda\rightarrow1}=0.01802$ as input, the shift of the transition point $\Delta\alpha_c=\alpha_c(\Lambda) - \alpha_{c,\Lambda\rightarrow1}$ is plotted in Fig.~\ref{f1}(b) as a function of the logarithm of the discretization factor $\ln \Lambda$ on a log-log scale. A power-law increase of $\Delta\alpha_c$ is found to provide a good fitting to the numerical data, and the value of the slope yields $d-1/\nu=1.79(2)$ from the scaling arguments and the relation $\ln\Lambda=-(\ln\omega_{\rm min})/M \sim 1/M$ where $M$ is equivalent to the length of the Wilson chain. By that we mean the Hamiltonian of SBM can be exactly mapped to the one describing a bosonic chain, i.e., Wilson chain, with nearest-neighbour interactions via a canonical transformation \cite{chin10,zho14}. In addition, quantum phase transition for the system with short-range interactions in $d$ spatial dimensions is generally believed to be equivalent to the classical transition in $d+1$ dimensions under the quantum-classical mapping. Thus taking effective spatial dimension $d_{\rm eff}=1+1$ by assumption, one calculates the correlation length exponent $1/\nu=0.21(2)$, in agreement with the mean-field prediction $1/\nu_{MF}=s=0.2$.

On the other hand, the dependence of the critical coupling $\alpha_c$ on the lowest frequency $\omega_{\rm min}$ is also investigated, and the results are shown in the inset where the discretization factor $\Lambda=2.0$ is set. Note the correlation length is given by an inverse energy scale $\xi = 1/\omega^{*}$ where $\omega^{*}$ is the frequency above which the quantum critical behavior is established. Therefore, a power-law relation $\Delta\alpha_c \propto L^{-1/\nu}=\omega_{\rm min}^{\rm 1/\nu}$ can be derived from the finite-size scaling analysis in frequency space. The exponent $1/\nu=0.22(1)$ is estimated from the slope of the curve in the inset, well consistent with the earlier one $0.21(2)$.

\subsubsection{Bath-related observables}
Critical behaviors of quantum entanglement and correlation within the sub-Ohmic bath are investigated in Fig.~\ref{f2} with the summations of the von-Neumann entropy $S_{\rm b}$, mutual information $\rm I_{\rm b}$, linear entropy $S_{\rm L,b}$, and quantum discord $D_{\rm b}$ on a linear scale. Obviously, they reach a sharp peak right at the phase transition, pointing to a cusp-like singularity. By normalizing the peak value to unity, all the curves have similar shapes, suggesting they obey the same scaling law in the sub-Ohmic regime. The transition point $\alpha_c=0.01803(1)$ is determined by the peak position which is indicated by the vertical dash-dotted line. It is in good agreement with the extrapolation result $\alpha_{c,\Lambda\rightarrow1}=0.01802$ in Fig.~\ref{f1}, showing $\Lambda=1.05$ is sufficient small for the convergence in the continuum limit. Besides, the accuracy of the variational result on the critical coupling is significantly improved, in comparison with previous ones $0.0168$ \cite{chi11} and $0.02014$ \cite{he18}. Moreover, it agrees well with numerical results $0.0185(4),0.0175(2)$, and $0.0179(5)$ obtained from the numerical renormalization group (NRG) \cite{bul03}, quantum Monte Carlo (QMC) \cite{win09}, and variational matrix product \cite{fre13}, respectively, thereby lending further support to the validity of variational calculations in this work.

\begin{figure*}[htb]
\epsfysize=7cm \epsfclipoff \fboxsep=0pt
\setlength{\unitlength}{1.cm}
\begin{picture}(10,7)(0,0)
\put(-3.0, 0.0){{\epsffile{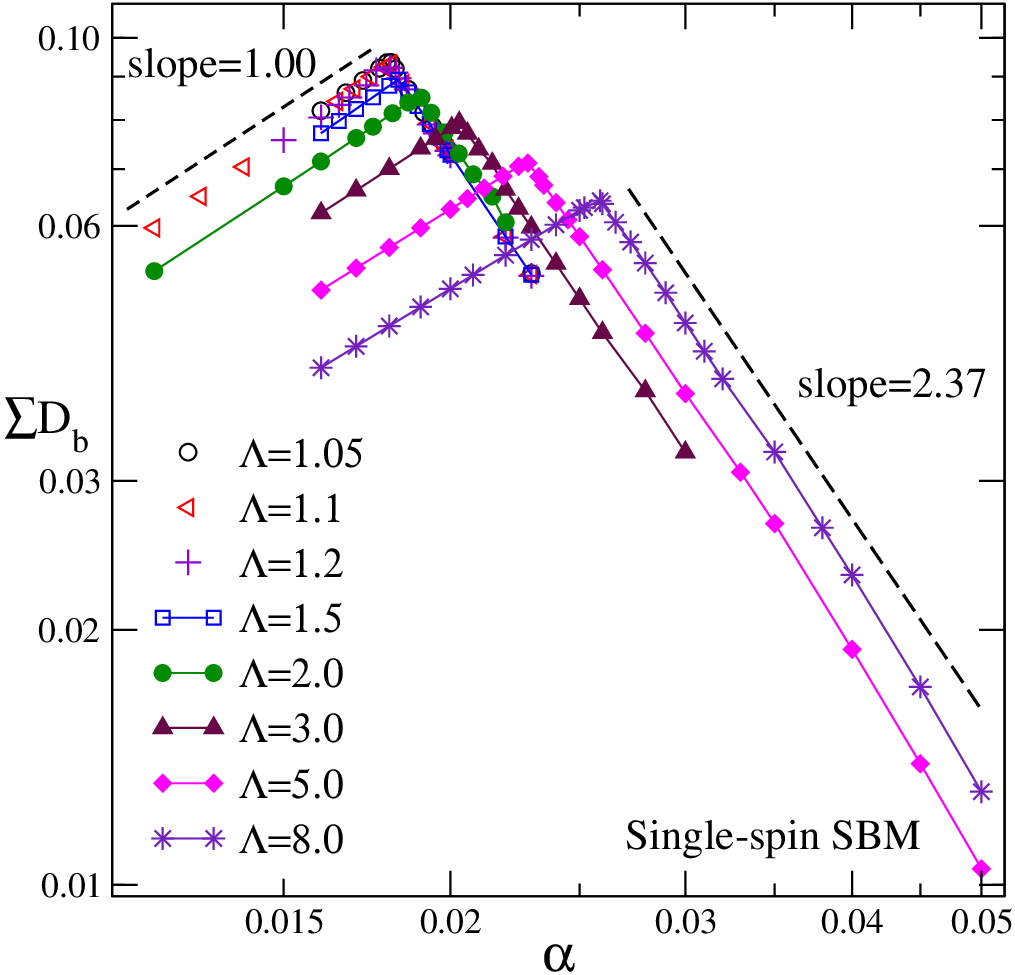}}}\epsfysize=7cm
\put(5.2, 0.0){{\epsffile{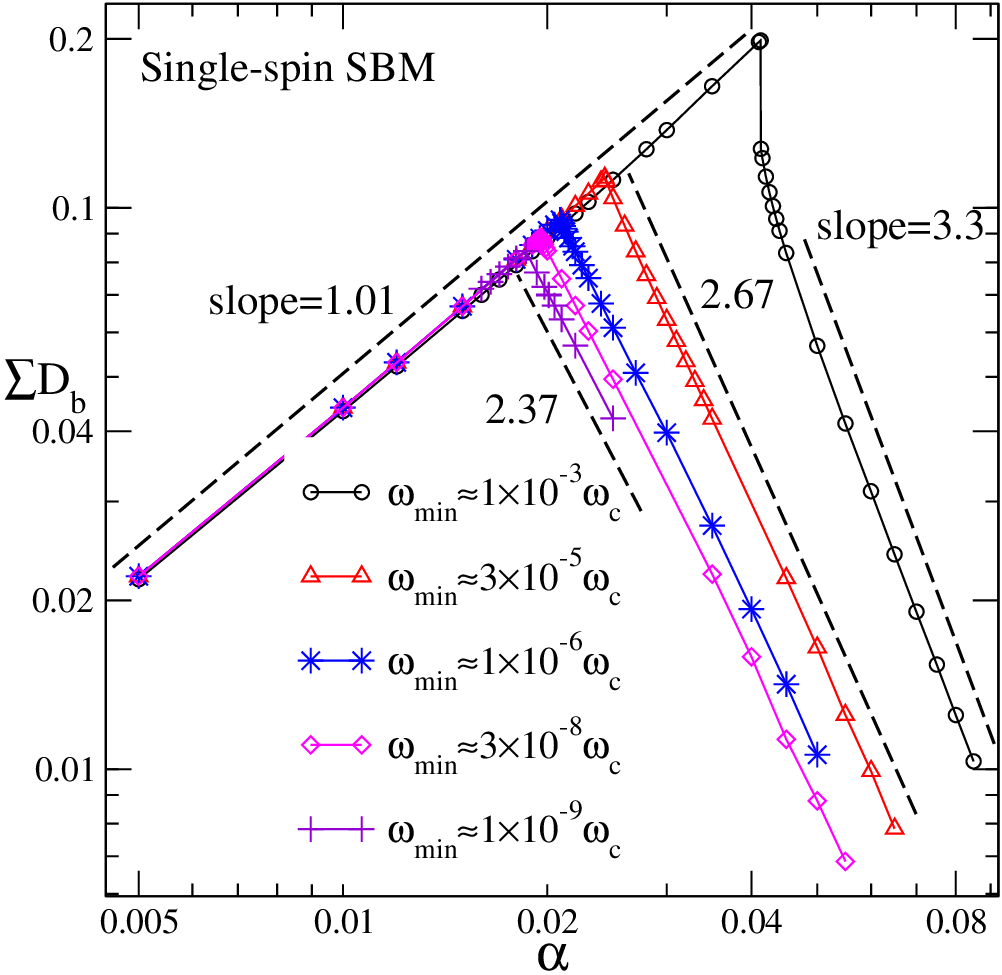}}}
\end{picture}

\hspace{0.0cm}\footnotesize{(a)}\hspace{8.0cm}\footnotesize{(b)}
\caption{ The summation of the quantum discord, $\sum{D_{\rm b}}$, is shown as a function of the coupling $\alpha$ for different discretization factors $\Lambda$ at $\omega_{\rm min}\approx 10^{-10}\omega_c$ in (a) and low-energy cutoff $\omega_{\rm min}$ at $\Lambda = 2.0$ in (b) on a log-log scale. Other parameters $s=0.2, \Delta/\omega_c=0.1$, and $N=8$ are set. In both (a) and (b), dashed lines represent power-law fits.
} \label{f3}
\end{figure*}

Using the summation of the quantum discord $\sum D_{\rm b}$ over the frequency $\omega_k$ as a representative indicator, one investigates the influence of the discretization factor $\Lambda$ in Fig.~\ref{f3}(a). Similar with that in Fig.~\ref{f1}(a), the critical coupling $\alpha_c$ determined by the peak position of $\sum D_{\rm b}$ is shifted left with the decreasing discretization factor $\Lambda$. While the slopes $1.00(2)$ and $2.37(3)$ in two sides are almost unchanged, showing that critical exponents are robust. Noting $\ln \Lambda \sim 1/M$, one concludes that the peak value of $\sum D_{\rm b}$ increases slightly with the environmental size $M$, confirming the general assumption that the quantum correlation has a singularity at the transition point.  Moreover, the dependence on the low-energy cutoff $\omega_{\rm min}$ is also demonstrated in Fig.~\ref{f3}(b). Interestingly, all the curves of $\sum D_{\rm b}$ for different $\omega_{\rm min} \approx 10^{-9}\sim 10^{-3}$ almost overlap with each other in the delocalized phase, indicating quantum discord $D_{\rm b}$ in the low-frequency regime is negligible. In the localized phase, however, the decay exponent is changed slightly from $3.3(1)$ to $2.37(4)$ with the decreasing $\omega_{\rm min}$, different from that in Fig.~\ref{f3}(a). It indicates the critical behavior of $\sum D_{\rm b}$ is dependent on $\omega_{\rm min}$ rather than $\Lambda$. The slope approaches the asymptotic value $2.37$ at $\omega_{\rm min} \approx 10^{-9}$, confirming that the value of the parameter $\omega_{\rm min}$ used in this work is already sufficiently small.

\subsubsection{Shallow sub-Ohmic regime}
As mentioned before, quantum phase transition in the deep sub-Ohmic regime $s<0.5$ is mean-field like where the critical exponent obeys $1/\nu=s$. In the shallow sub-Ohmic regime $s>0.5$, however, non-trivial critical behaviors have been found in both the numerical work and analytical analysis \cite{voj05,ort10,win09}. The hyperscaling relations hold, and the exponent $\nu$ is expected to diverge as $1/\sqrt{2(1-s)}$ near the ohmic point $s=1$. In order to demonstrate the validity of our scaling analysis and NVM in such non-mean-filed regime, additional simulations with the spectral exponent $s=0.7$ are performed for different values of $\Lambda$ and $\omega_{\rm min}$.

\begin{figure*}[htb]
\centering
\epsfysize=7cm \epsfclipoff \fboxsep=0pt
\setlength{\unitlength}{1.cm}
\begin{picture}(10,7)(0,0)
\put(-3.0, 0.0){{\epsffile{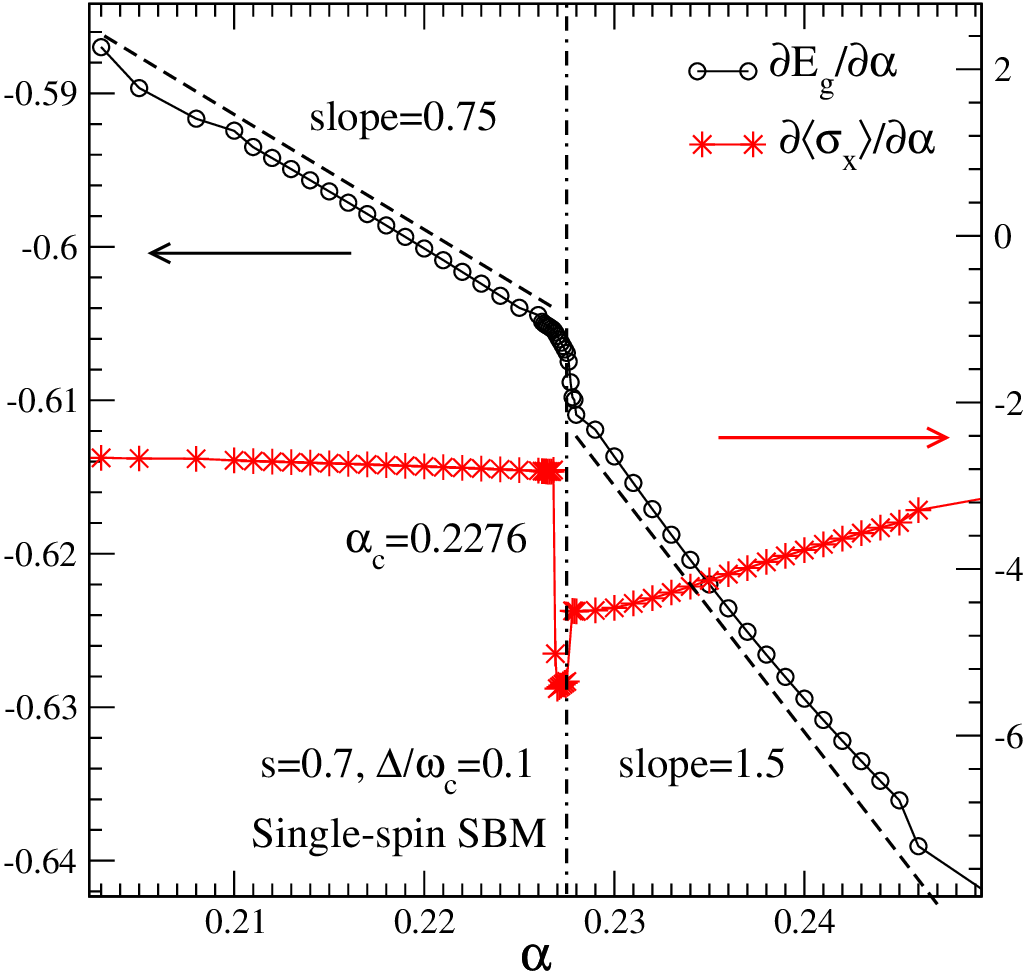}}}\epsfysize=7cm
\put(5.2, 0.0){{\epsffile{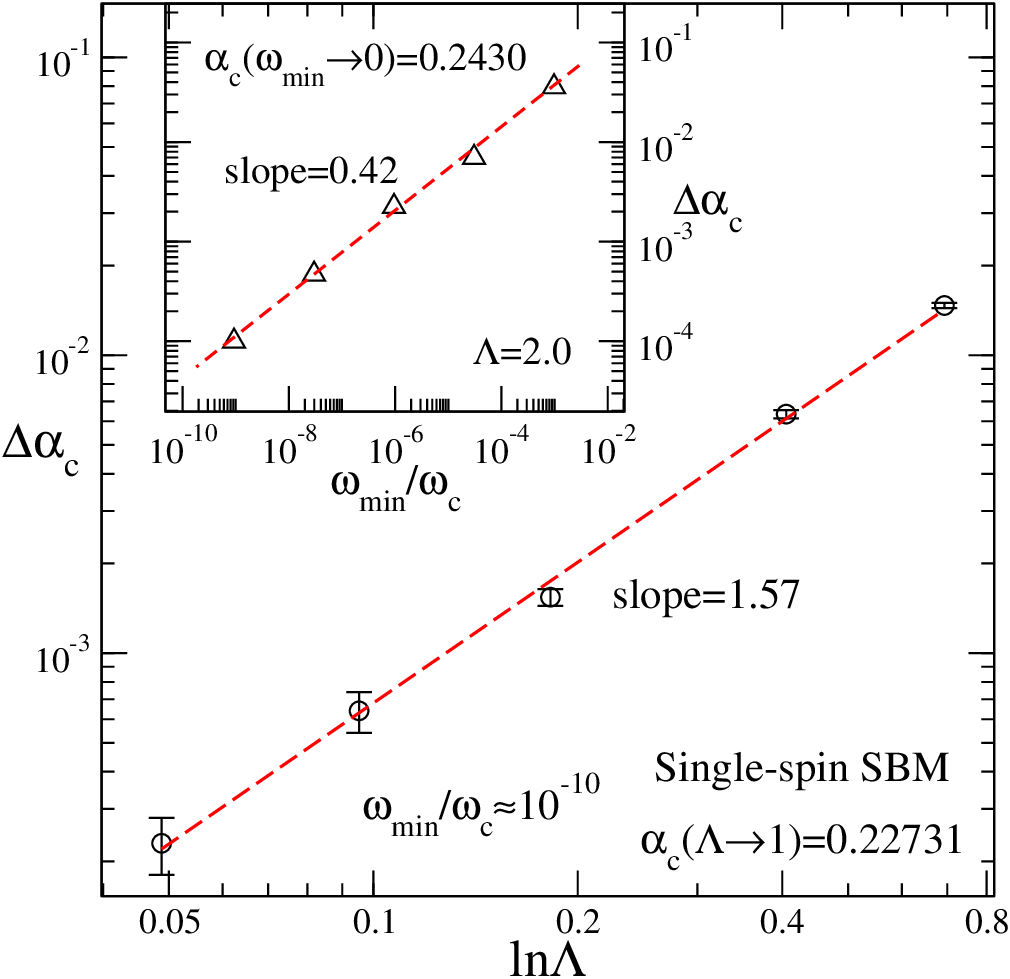}}}
\end{picture}

\hspace{0.0cm}\footnotesize{(a)}\hspace{8.0cm}\footnotesize{(b)}
\caption{(a) In the shallow sub-Ohimc SBM ($s=0.7$), the first derivatives of the ground-state energy $E_g$ as well as the spin coherence $\langle\sigma_x\rangle$ are displayed as a function of the coupling strength $\alpha$ on a linear scale. The tunneling constant $\Delta=0.1\omega_c$, discretization factor $\Lambda=1.05$, and numbers of coherent-superposition states and effective bath modes $N=8, M=430$ are set. The left and right arrows indicate the Y coordinates for $\partial E_g/\partial \alpha$ and $\partial \langle\sigma_x\rangle/\partial \alpha$, respectively. (b) The shift of the transition point $\Delta\alpha_c$ with respect to $\ln \Lambda$ and $\omega_{\rm min}$ is plotted with open circles and open triangles (in the inset), respectively. Dashed lines represent power-law fits.}
\label{add_f1}
\end{figure*}

Taking $\Lambda=1.05$ as an example, the first derivative of the ground-state energy, $\partial E_g/\partial \alpha$, is plotted versus the coupling $\alpha$ in Fig.~\ref{add_f1}(a). Similar to that in Fig.~\ref{f1}, high-order singularity is obtained at the transition point $\alpha_c=0.2276(1)$, though the slope difference between two phases ($1.5-0.75=0.75$) becomes far less, inferring that the transition is weakened in the shallow sub-Ohmic regime. In addition, the first derivative of the spin coherence, $\partial\langle\sigma_x\rangle/\partial \alpha$, is also presented with stars, and a tiny discontinuity is detected at the critical point, again supporting the transition is of second order.

In Fig.~\ref{add_f1}(b), the asymptotic behavior of $\Delta \alpha_c$ is carefully examined. By taking into account the mass flow corrections, the transition point $\alpha_c=0.2430$ is estimated in the inset at $\Lambda=2$, in agreement with QMC result $0.241(2)$ \cite{win09}. The asymptotic value of $\alpha_c$ is then refined to be $0.22731$ in the continuum limit $\Lambda \rightarrow 1$, corresponding to a high dense sub-Ohmic bath. Moreover, the exponent value $1/\nu=0.43(3)$ is determined, well consistent with the QMC result $0.45(5)$, but far away from the predictions $1/\nu=s=0.7$ and $\sqrt{2(1-s)}\approx0.775$. Quantum-to-classical correspondence gives that the SBM quantum transition is in the same universality class as the classical Ising chain with long-range interactions decaying as $1/r^{1+s}$ \cite{voj05}. Interestingly, our result agrees well with  $1/\nu=1/2+1/3\epsilon-2.628\epsilon^2+\mathcal {O}(\epsilon^3)\approx 0.461$ with $\epsilon=s-0.5$ which can be read off from the hyperscaling relation $\gamma=(2-\eta)\nu$ and two-loop renormalization-group results on the exponents $\gamma$ and $\eta$ in the Ising chain \cite{fis72}.

\subsection{Kosterlitz-Thouless phase transitions}
Besides sub-Ohmic quantum transitions, quantum phase transitions in the Ohmic SBM ($s=1$) which are of Kosterlitz-Thouless type are also studied in the weak tunneling  and continuous limits $\Delta \rightarrow 0, \Lambda \rightarrow 1$. Considering the constraint available computational resources, the tunneling amplitude $\Delta=0.01$ and discretization factor $\Lambda=1.01$ are set as a demonstration. Convergence check shows that the numbers of coherent-superposition states $N=6$ and of effective bath modes $M=1000$, are sufficiently large for Ohmic phase transitions.

\begin{figure*}[htb]
\centering
\epsfysize=8cm \epsfclipoff \fboxsep=0pt
\setlength{\unitlength}{1.cm}
\begin{picture}(9,8)(0,0)
\put(0.0,0.0){{\epsffile{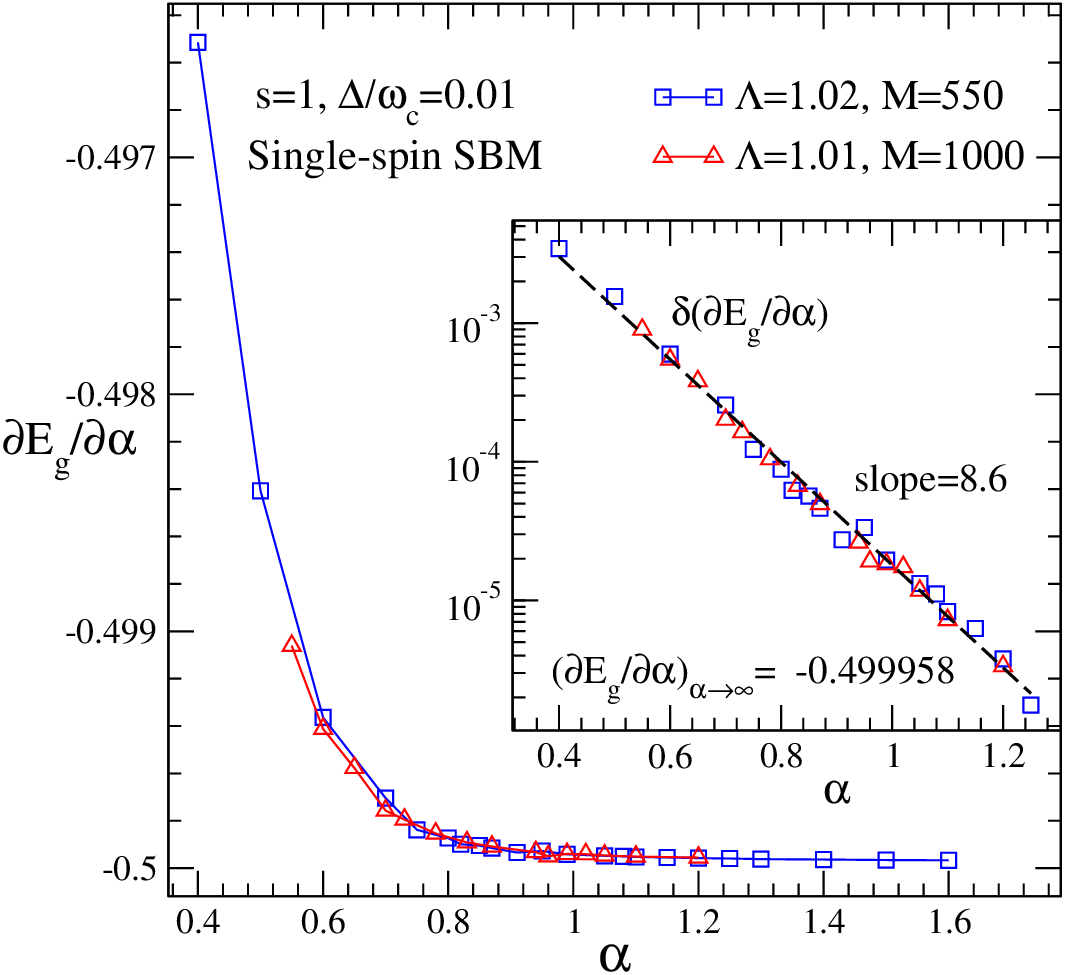}}}
\end{picture}
\caption{ In the Ohmic SBM with $s=1$, the ground-state energy derivative $\partial E_g/\partial \alpha$ is displayed for the logarithmic discretization factors $\Lambda=1.01$ and $1.02$. The tunneling constant $\Delta/\omega_c=0.01$ and the number of coherent-superposition states $N=6$ are set. In the inset, the shift $\delta(\partial E_g/\partial \alpha) = (\partial E_g/\partial \alpha)- (\partial E_g/\partial \alpha)|_{\alpha \rightarrow \infty}$ is plotted with open triangles ($\Lambda=1.01$) and open squares ($\Lambda=1.02$), respectively, on a linear-log scale. Dashed line represents an exponential fit.  }
\label{f4}
\end{figure*}

The first derivative of the ground-state energy $\partial E_g/\partial \alpha$ is plotted in Fig.~\ref{f4} for the discretization factors $\Lambda=1.01$ on a linear scale. For comparison, numerical simulations with $\Lambda=1.02$ are also performed . The overlap of two curves indicates that the effect of discretization is already negligibly small. With the increasing coupling $\alpha$, the ground-state energy derivative $\partial E_g/\partial \alpha$ exhibits a smooth decay, tending to a converging value around $-0.5$. Taking the asymptotic value $-0.499958$ as input, an exponential decay of the shift $\delta(\partial E_g/\partial \alpha)$ is found with the slope $8.6$, as shown in the inset. It suggests there is no discontinuity in derivatives of the ground-state energy $E_g$ of any order, supporting the quantum phase transition is of the Kosterlitz-Thouless type.

\subsubsection{Bath criticality in single-spin SBM}

\begin{figure*}[htb]
\centering
\epsfysize=8cm \epsfclipoff \fboxsep=0pt
\setlength{\unitlength}{1.cm}
\begin{picture}(9,8)(0,0)
\put(0.0,0.0){{\epsffile{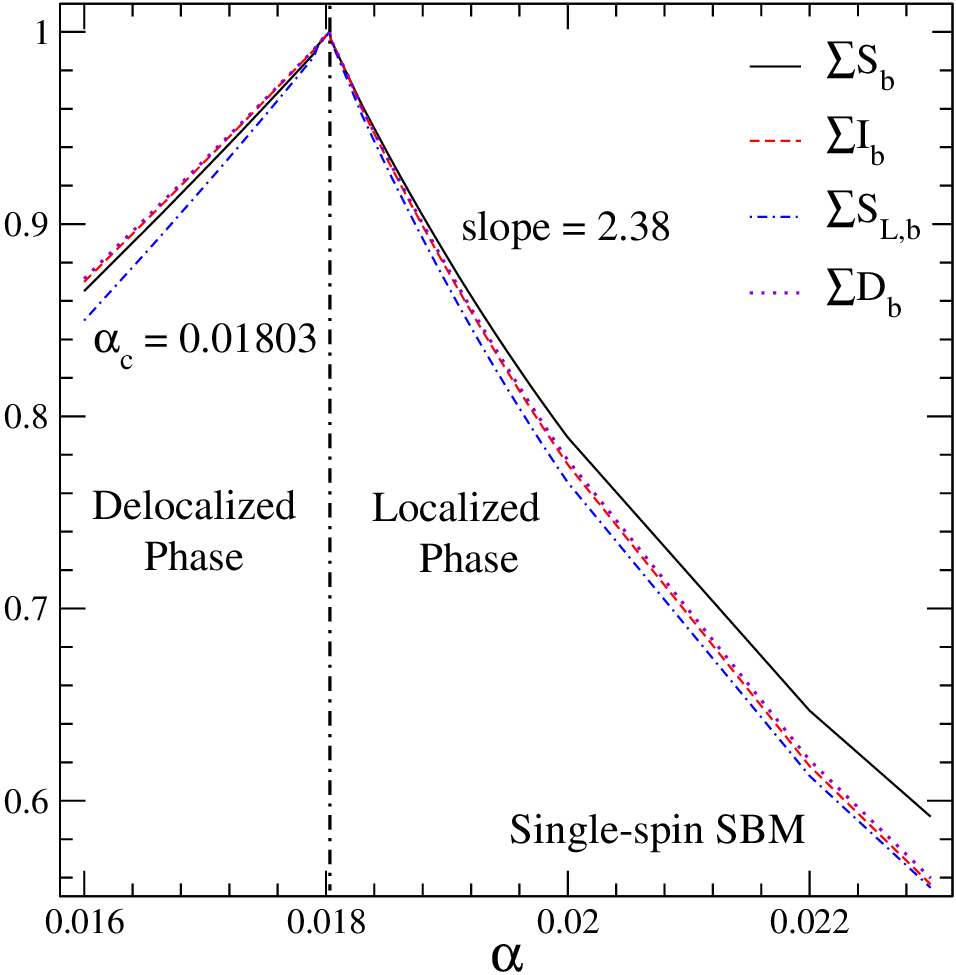}}}
\end{picture}
\caption{ The summations of the von-Neumann entropy $S_{\rm b}$, mutual information $\rm I_{\rm b}$, linear entropy $S_{\rm L,b}$, and quantum discord $D_{\rm b}$ are plotted with the solid, dashed, dash-dotted and dotted lines, respectively, on a linear scale. The discretization factor $\Lambda=1.05$ and the numbers of coherent-superposition states and effective bath modes $N=8, M=430$ are set. For clarity, all the values of the peaks are scaled to the unit. The transition point $\alpha_c=0.01803(1)$ is then located by the vertical line.   }
\label{f2}
\end{figure*}

\begin{figure*}[b]
\epsfysize=7cm \epsfclipoff \fboxsep=0pt
\setlength{\unitlength}{1.cm}
\begin{picture}(10,7)(0,0)
\put(-3.0, 0.0){{{\tiny \label{key}}\epsffile{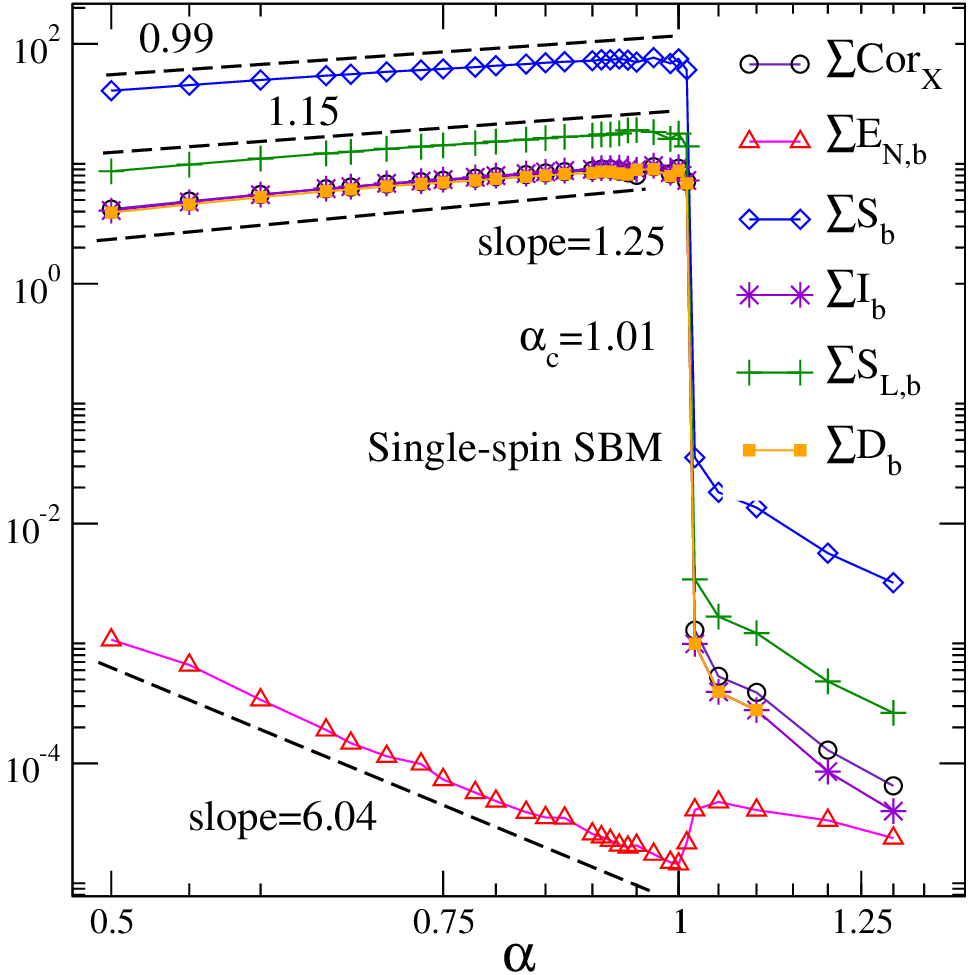}}}\epsfysize=7cm
\put(5.2, 0.0){{\epsffile{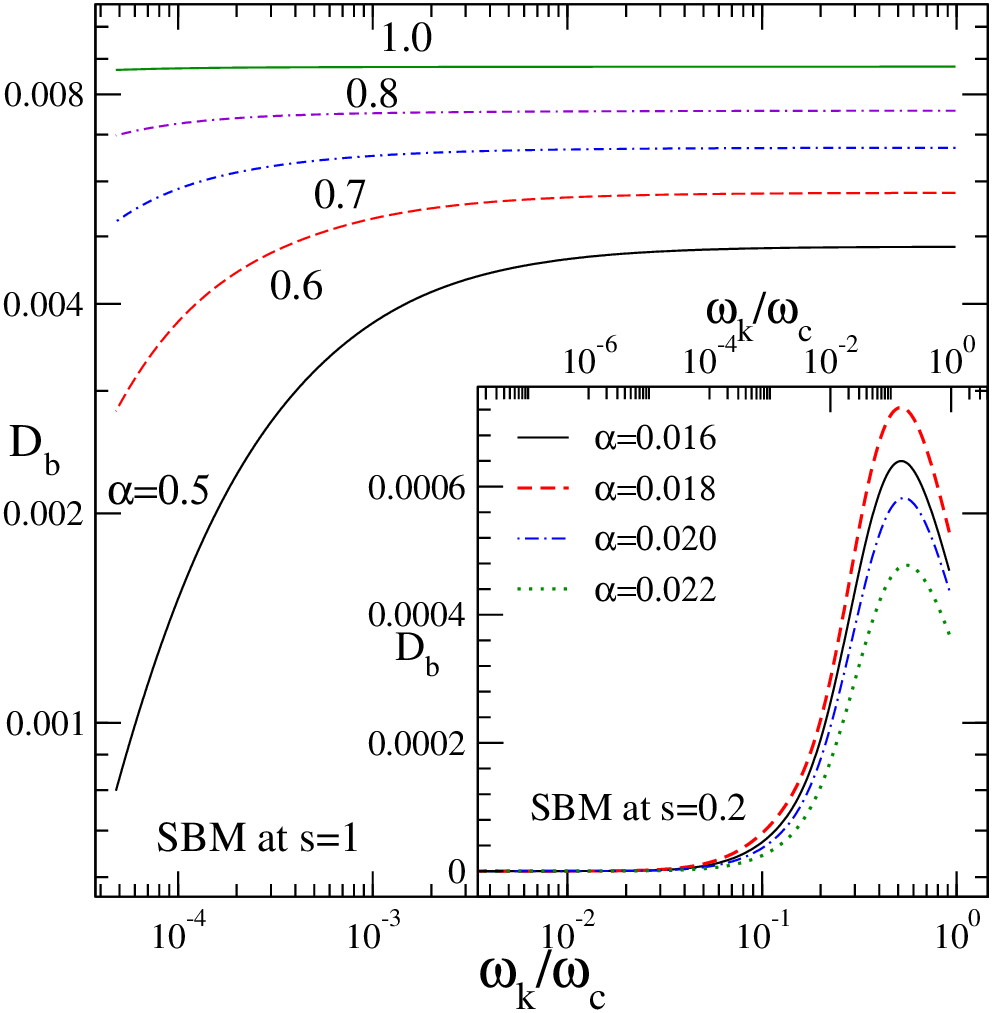}}}
\end{picture}

\hspace{0.0cm}\footnotesize{(a)}\hspace{8.0cm}\footnotesize{(b)}
\caption{(a) The summations of the correlation function $\rm Cor_X$, logarithmic negativity $E_{\rm N,b}$, von-Neumann entropy $S_{\rm b}$, mutual information $I_{\rm b}$, linear entropy $S_{\rm L,b}$, and quantum discord $D_{\rm b}$ are plotted versus the coupling strength $\alpha$ on a log-log scale. The dashed lines show power-law fits. Other parameters $\Delta/\omega_c=0.01, s=1, \Lambda=1.01, N=6$, and $M=1000$ are set. (b) The quantum discord $D_{\rm b}$ is plotted as a function of the bosonic frequency $\omega_k$ for different couplings.  For comparison,  the quantum discord $D_{\rm b}(\omega_k)$ in the sub-Ohmic regime $s=0.2$ is also given in the inset.
} \label{f5}
\end{figure*}

In Fig.\ref{f5}(a), the summations of the correlation function $\rm Cor_X$, logarithmic negativity $E_{\rm N,b}$, von-Neumann entropy $S_{\rm b}$, mutual information $I_{\rm b}$, linear entropy $S_{\rm L,b}$, and quantum discord $D_{\rm b}$ are plotted on a log-log scale for the quantum entanglement and  correlation within the Ohmic bath. Different from the cusp-like singularity in Fig.~\ref{f2}, the discontinuities analogous to the universal jump of the superfluid density in the XY model are found in all curves, again pointing to the emergence of Kosterlitz-Thouless phase transition. The critical value of the coupling $\alpha_c=1.01(1)$ is then estimated for $\Delta=0.01$, in good agreement with the prediction $\alpha_c=1+\mathcal {O}(\Delta/ \omega_c)$ \cite{hur08}. Furthermore, the asymptotic value $\alpha_c=1.0053$ is measured by the extrapolation $\omega_{\rm min} \rightarrow 0$, and the linear coefficient $(\alpha_c-1)\omega_c/\Delta=5.3$ is excellent consistent with that in Ref.~\cite{fil20}. It points out that our variational method is as powerful and efficient as QMC and variational Feynman in providing an accurate description of the physical features of SBM.

In the delocalized phase $\alpha < \alpha_c$, the summations of the von-Neumann entropy $S_{\rm b}$ and its linear term $S_{\rm L,b}$ increase with the coupling $\alpha$ as a power-law form with the slope close to $1$, pointing to the linear dependence. A good coincidence is found for the curves of the correlation function $\rm Cor_X$, mutual information $I_{\rm b}$, and quantum discord $D_{\rm b}$, indicating the bath embodies pure quantum effect. They behave similarly to the entropy, though the slope $1.25(2)$ is slightly larger than $1$. In contrast, the summation of the logarithmic negativity $E_{\rm N,b}$ exhibits a power-law decay with a larger exponent $6.04(8)$, suggesting the average bipartite entanglement is rapidly erased by the decoherence effect from the environment. The opposite trend and vanishing value of $\sum E_{\rm N,b}$ show that the ground state of the Ohmic bath in the delocalized phase is separable  mixed, rather than pure entangled.

Subsequently, the frequency dependence is examined, typified by the quantum discord $D_{\rm b}$. In Fig.~\ref{f5}(b), $D_{\rm b}(\omega_k)$ is displayed for different couplings $\alpha =0.5, 0.6,0.7,0.8$, and $1.0$ on a log-log scale. It increases monotonously with the frequency $\omega_k$, and quickly stabilizes at a constant which is coupling dependent. As the coupling $\alpha$ increases, the growth curve flattens out gradually, and tends to be $\omega_k$-independent at the transition point. For comparison, the quantum discord $D_{\rm b}(\omega_k)$ in the sub-Ohmic regime $s=0.2$ is also given in the inset. Different from the Ohmic case, all the curves of $D_{\rm b}(\omega_k)$ exhibit sharp peaks in the high-frequency region. It is worth noting that the peak value reaches a maximum at the transition $\alpha_c=0.018$, while the position of the peak remains practically unchanged. Therefore, two distinct scaling behaviors of the bosonic bath can be found for the Ohmic and sub-Ohmic transitions, lending further support to the claim that they belong to different universality classes.

\begin{figure*}[htb]
\centering
\epsfysize=7cm \epsfclipoff \fboxsep=0pt
\setlength{\unitlength}{1.cm}
\begin{picture}(10,7)(0,0)
\put(-3.0, 0.0){{\epsffile{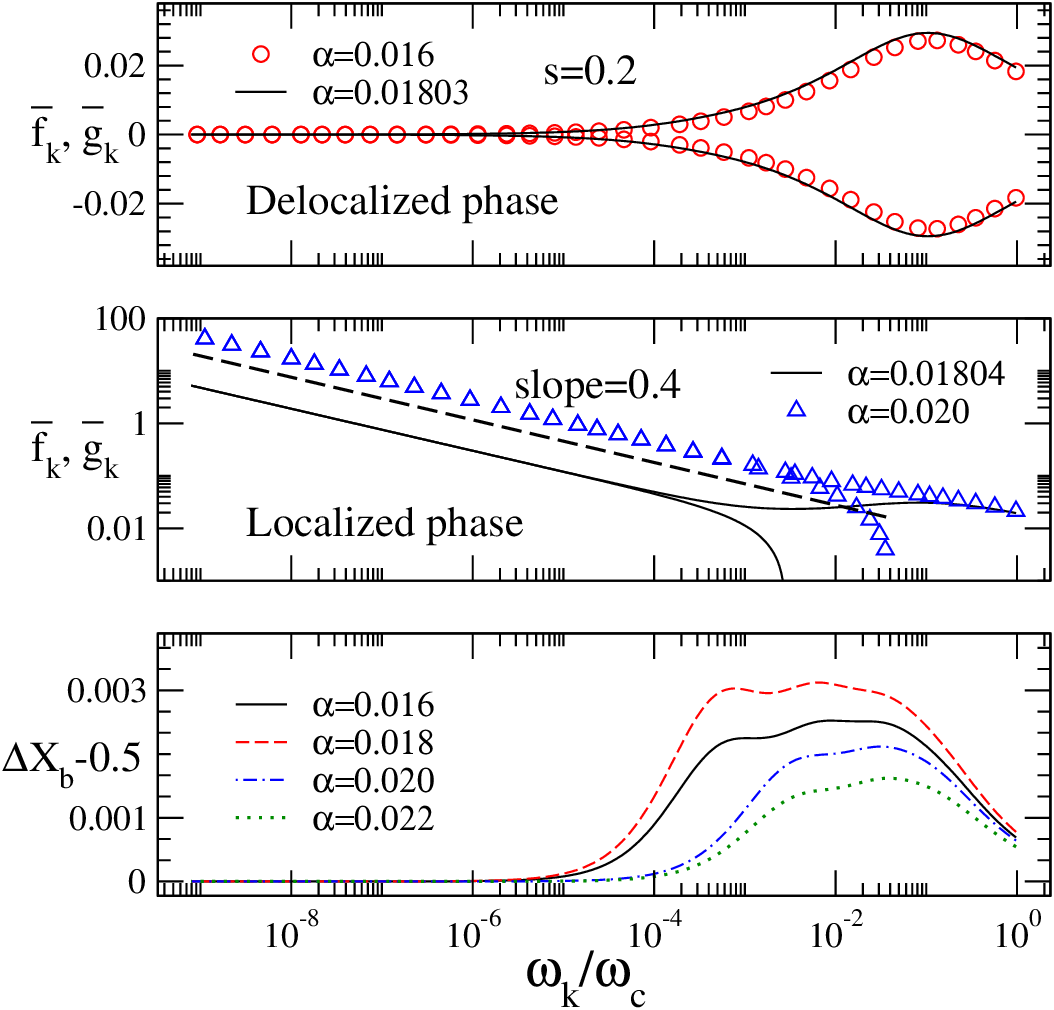}}}\epsfysize=7cm
\put(5.2, 0.0){{\epsffile{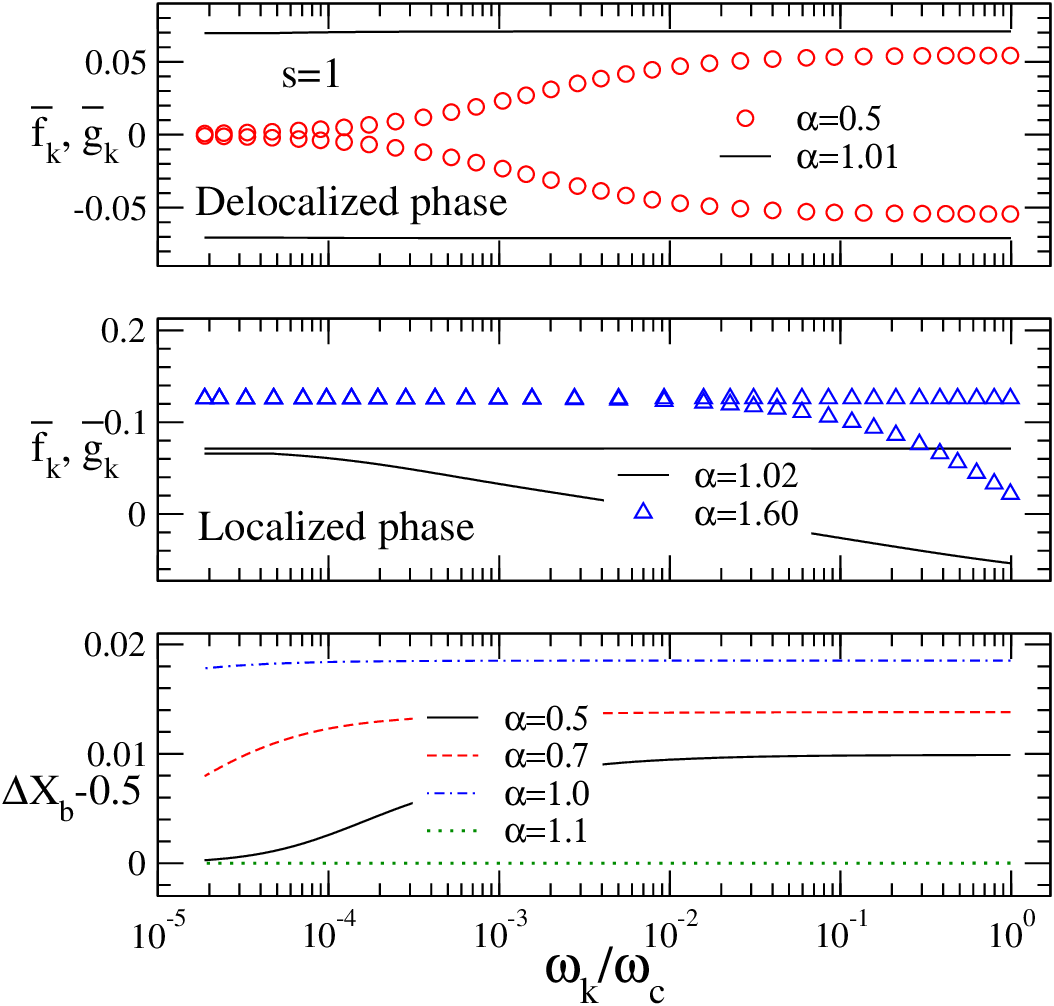}}}
\end{picture}

\hspace{0.0cm}\footnotesize{(a)}\hspace{8.0cm}\footnotesize{(b)}
\caption{The frequency-dependent average displacement coefficients $\overline f_k$  and $\overline g_k$ as well as the quantum fluctuation $\Delta X_{b}-1/2$ in the position space are plotted for different couplings $\alpha$ in (a) and (b), corresponding to the sub-Ohmic SBM ($s=0.2$) and Ohmic SBM ($s=1$), respectively. Dashed line in the middle panel of the subfigure (a) represents an power-law fit. }
\label{add_f2}
\end{figure*}

For better understanding the nature of ground states in the sub-Ohmic and Ohmic cases, the structure of the environmental wave function characterized by average displacement coefficients defined as $\overline f_k=\langle \Psi_1|(b_k+b_k^{\dag})(1+\sigma_z)|\Psi_1\rangle/2$ and $\overline g_k=\langle\Psi_1|(b_k+b_k^{\dag}) (1-\sigma_z)|\Psi_1\rangle/2$, is demonstrated in Fig.~\ref{add_f2}. For the sub-Ohmic SBM at $s=0.2$, a perfect antisymmetry relation $\overline f_k =- \overline g_k$ is found over the whole range of $\omega_k$ in the upper panel of the subfigure (a), supporting the usual assumption concerning the delocalized phase \cite{sil84,ber14}. A huge jump appears in the low-frequency value of the displacement coefficient when the coupling is changed from $\alpha=0.01803$ to $0.01804$, showing a sharp transition. In the middle panel, power-law decays of both $\overline f_k$ and $\overline g_k$ are observed with the slope $0.4$ for $\alpha > \alpha_c$ at low frequencies, confirming that they follow the same classical displacement $\lambda_k/(2 \omega_k) \sim \omega_k^{-(1-s)/2}=\omega_k^{-0.4}$, and thereby the antisymmetry is broken in the localized phase.

For the Ohmic SBM at $s=1$, the antisymmetry relation as well as the antipolaron which was proposed as an important concept in Ref.~\cite{ber14} is confirmed in our numerical work, indicating that the picture of the variational parameters in the delocalized phase is almost the same. Spontaneous breakdown of the antisymmetry is also found at the quantum phase transition, as shown in Fig.~\ref{add_f2}(b). While the deep Kondo regime where $\alpha \rightarrow 1$ and the localized phase were untouched in that work \cite{ber14}. Further analysis points out that the amount of the jump for the average displacement coefficient at $\omega_{\rm min}$ is four orders of magnitude smaller than the sub-Ohmic one. That is because the value of $\overline f_k$ or $\overline g_k$ is independent of $\omega_k$ in the localized phase, for the classical displacement $\lambda_k/(2\omega_k)=\rm constant$. Hence, it can be inferred that the bath modes with low and high frequencies may follow the same critical scaling in the Ohmic case, but behave differently in the sub-Ohmic regime.

Quantum fluctuations mainly caused by the effect of the antipolarons are investigated in the lower panels of Fig.~\ref{add_f2}, which can be measured by the departure from the single-coherent state, $\Delta X_{\rm b}-1/2$. In the sub-Ohmic regime, as expected, quantum fluctuations at low frequencies vanish in both the localized and delocalized phases. In contrast, the curve develops a plateau in the high-frequency region for any coupling $\alpha$, and the plateau value reaches the maximum at the transition point $\alpha_c=0.018$, similar with that in the inset of Fig.~\ref{f5}(b). It suggests that the emergences of kinks in Figs.~\ref{f2} and \ref{f3} are only related to the high-frequency bath modes. In the Ohmic case, the quantum fluctuation $\Delta X_{\rm b} - 1/2$ for $\alpha=0.5,0.7$, and $1.0$ grows with the frequency $\omega_k$, and approaches a $\alpha$-dependent constant value. For a larger coupling, e.g., $\alpha=1.1$, however, it vanishes $\Delta X_{\rm b} - 1/2=0$ over the whole frequency range, confirming that bath modes are independent of each other, and behave as a single-coherent state in the localized phase. The picture is quite different from that in the sub-Ohmic regime.

\subsubsection{Bath criticality in two-spin SBM}

\begin{figure*}[htb]
\epsfysize=7cm \epsfclipoff \fboxsep=0pt
\setlength{\unitlength}{1.cm}
\begin{picture}(10,7)(0,0)
\put(-3.0, 0.0){{\epsffile{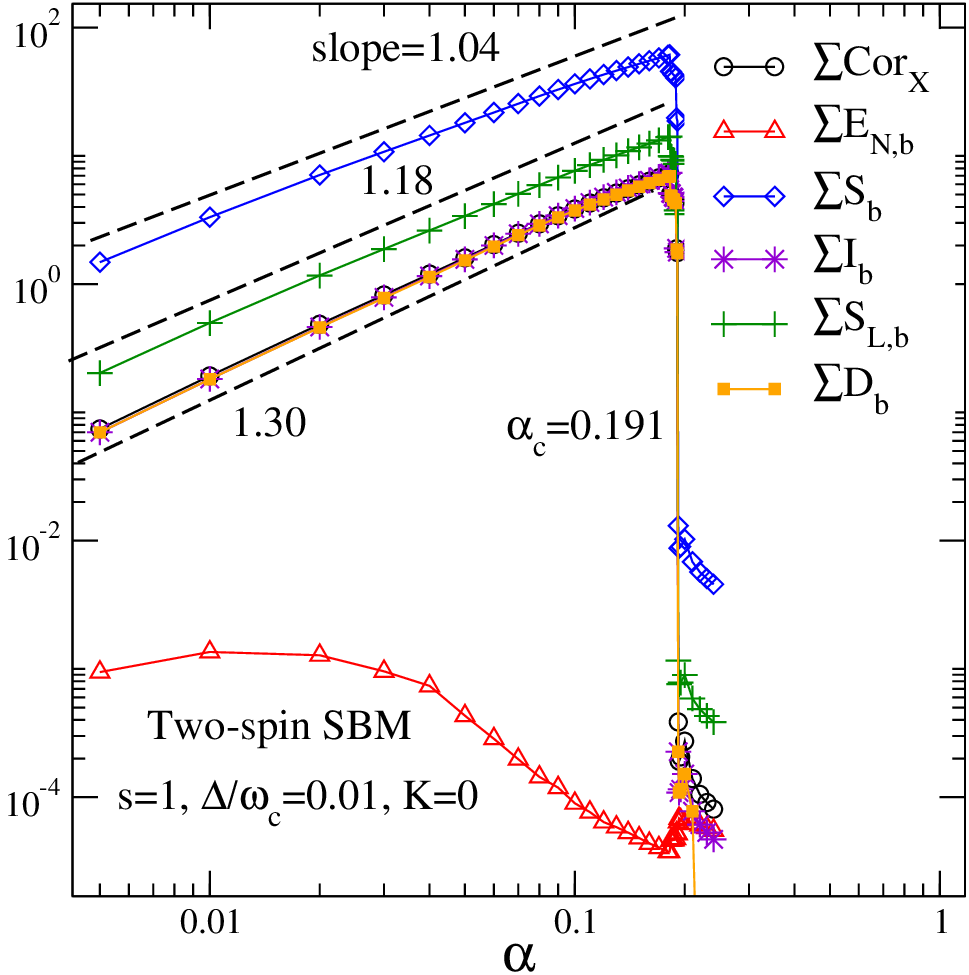}}}\epsfysize=7cm
\put(5.2, 0.0){{\epsffile{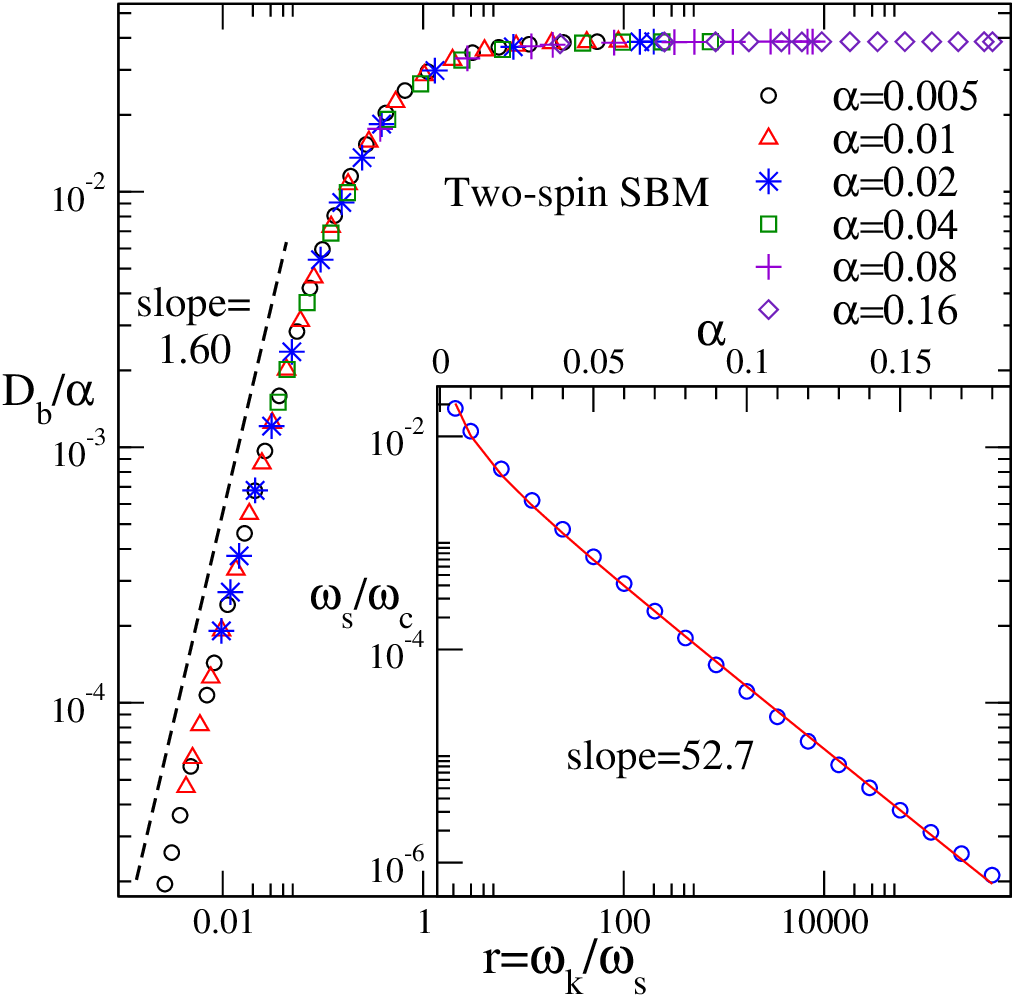}}}
\end{picture}

\hspace{0.0cm}\footnotesize{(a)}\hspace{8.0cm}\footnotesize{(b)}
\caption{(a) In the Ohmic bath $s=1$ of two-spin SBM with vanishing spin-spin interaction $K=0$, the summations of the correlation function $\rm Cor_X$, logarithmic negativity $\rm E_{\rm N,b}$, von-Neumann entropy $\rm S_b$, mutual information $\rm I_b$,  linear entropy $S_{\rm L,b}$, and quantum discord $\rm D_b$ are plotted on a log-log scale at $\Delta/\omega_c=0.01, \Lambda=1.01, N=6$ and $M=1000$. (b) The scaling function of the quantum discord $\widetilde{D}_b(r) =\rm D_{\rm b}/\alpha$ is displayed with respect to the ratio $r=\omega_k/\omega_s$ for different couplings. Inset shows the energy scale $\omega_s(\alpha)$ extracted from the data collapse. In both (a) and (b), dashed lines represent power-law fits, and solid line in the inset shows the fit with an exponential form.
} \label{f6}
\end{figure*}

Environmental entanglement, correlation, and entropy in two-spin SBM are also presented in Fig.~\ref{f6}(a), in the presence of the vanishing spin-spin coupling $K=0$ and weak tunneling $\Delta=0.01$. Similar to those in Fig.\ref{f5}(a), they follow power-law behaviors in the delocalized phase, though the values of the growth exponents are slightly larger. The transition point $\alpha_c=0.191(2)$ is then determined by sudden drops of the curves, much lower than earlier variational results $\alpha_c=0.5$ \cite{mcc10} and $0.31$ \cite{zho18}. It is because that the value of critical coupling can be well refined with the help of the high dense spectrum and broad frequency range $\omega_c/\omega_{\rm min} > 10^{5}$ \cite{zho21}. Moreover, it is in good agreement with the NRG one $\alpha_c=0.185$  and QMC one $0.210$ estimated from the linear extrapolation $\Delta \rightarrow 0.01$ of numerical results obtained in previous work \cite{ort10,win14}, further confirming the accuracy of our NVM calculations.

General scaling arguments for the continuous phase transition lead to the scaling form of the quantum discord,
\begin{equation}
\label{scale}
D_{\rm b}(\omega_k, \alpha) = \alpha^{\lambda} \widetilde{D}_b(\omega_k/\omega_s),
\end{equation}
where $\alpha^{\lambda}$ indicates the scaling dimension, $\widetilde{D}_b(r=\omega_k/\omega_s)$ represents the scale invariance of the quantum discord in the Ohmic bath, and $\omega_s=1/\xi$ denotes the inverse of the correlation length.

The scaling function of the quantum discord $\widetilde{D}_b(r) =\rm D_{\rm b}\alpha^{-\lambda}$ defined in Eq.~(\ref{scale}) is shown in Fig.~\ref{f6}(b) with respect to the ratio $r=\omega_k/\omega_s$. With the exponent $\lambda=1$ and energy scale $\omega_s(\alpha)$ as inputs, all data of different coupling $\alpha$ nicely collapse to a single curve, fully confirming the scaling form of the quantum discord $D_{\rm b}$. Clearly, $\widetilde{D}_b(r) \rightarrow {\rm const}$ when $r \rightarrow \infty$, and $\widetilde{D}_b(r) \sim r^{1.60}$ when $r \rightarrow 0$, suggesting it is a non-analytic function. In the inset, the energy scale $\omega_s$ extracted from the data collapse decays exponentially with the coupling $\alpha$. Solid line provides a good fit to the numerical data with a power-law correction to scaling, yielding the exponent value $52.7(4)$. It is perfectly consistent with the result $52.1(5)$ measured from the magnetization $\langle\sigma_z\rangle$ under a tiny bias $10^{-5}$, and almost twice as large as that from the renormalized tunneling $\Delta_r$ \cite{zho18}. Therefore, it is concluded that exponentially divergent correlation length $1/\omega_s$ plays an essential role in critical behaviors of two-spin SBM.

\subsubsection{Spin criticality in two-spin SBM}

For comparison, the correlation, entanglement, and entropy of two spins are also investigated in this work. Firstly, the von Neumann entropy $S_{\rm{v-N}}$ that characterizes the entanglement between the spin system and its surrounding bath is introduced, $S_{\rm{v-N}}=-\rm Tr[\rho_{s}log_2\rho_s]$, where $\rho_s=\rm Tr_b[\rho_{sb}]$ is a reduced system density matrix given by tracing the total (system + bath) density operator $\rho_{sb}$ over the bosonic bath. The linear entropy of the system is then calculated as $S_L=\rm 1-Tr[\rho_{s}\rho_{s}]$. With the reduced density matrix $\rho_s$ at hand, the quantum entanglement between two spins can be measured by the concurrence,
\begin{equation}
\label{concurrence}
C(\rho_s)=\rm max\{\lambda_1-\lambda_2-\lambda_3-\lambda_4,0\},
\end{equation}
where $\lambda_i$ ($i=1,2,3,$ or $4$) represents a square root of the eigenvalues of the matrix $\rho_s\widetilde\rho_s$ arranged in a descending
order, and $\widetilde\rho_s=(\sigma_y\otimes\sigma_y)\rho_s^*(\sigma_y\otimes\sigma_y)$. The entanglement of the formation
is then calculated,
\begin{equation}
S_{\rm E}(C)=h\left(\frac{1+\sqrt{1-C^2}}{2}\right),
\end{equation}
with the function $h(x)=-x\log_2 x-(1-x)\log_2 (1-x)$.

In addition, the spin-spin correlation function $\rm Cor=\langle\sigma_{z1}\sigma_{z2}\rangle - \langle\sigma_{z1}\rangle \langle \sigma_{z2}\rangle $ and mutual information $I=S_{\rm{v-N}}(\rho_{s1})+S_{\rm{v-N}}(\rho_{s2})-S_{\rm{v-N}}(\rho_s)$ are investigated for the correlation in the spin system, where the subscripts $1$ and $2$ denote the ranks of the spins. The negativity is a measure of quantum entanglement derived from the PPT criterion for separability, which is an entanglement monotone.  The negativity of a subsystem can be defined as $N(\rho_s)=(||\rho_s^{\rm T}||_1 - 1)/2$, where $\rho_s^{\rm T}$ is the partial transpose of the reduced system density $\rho_s$ with respect to the spin $1$, and $||\hat{X}||_1=\text{Tr}|\hat{X}|=\text{Tr}\sqrt{\hat{X}^{\dagger}\hat{X}}$ is the trace norm or the sum of the singular values of the operator $\hat{X}$. The logarithmic negativity is then calculated as $E_{\rm N}=\log_2(2N+1)$. It is believed to be an easily computable entanglement measurement and an upper bound to the distillable entanglement \cite{hor09}.

Finally, the quantum discord reflecting the nonclassical part of the total correlation is calculated as $D(\rho_{s2|1}) =I(\rho_s)  - C(\rho_{s2}:\rho_{s1})$ where $I$ denotes mutual information between two spins $1$ and $2$, and $C$ represents the classical correlation,
\begin{align}
& C\left(  \rho_{s2}:\rho_{s1}\right)   \equiv \sup_{\left \{ \Pi_{j1}\right \} }I\left(  \rho_s|\left \{  \Pi_{j1}\right \}  \right) \\ \nonumber
& = \sup_{\left \{  \Pi_{j1}\right \} }\left[ S_{\rm v-N}\left( \rho_{s2}\right)  -\sum_{j}p_{j}S_{\rm v-N}\left( \rho_{j2}\right)\right],
\end{align}
given by a certain projection measurement $\left \{
\Pi_{j1}\right \}  $ on the spin $1$ with
\begin{equation}
p_{j}=\text{Tr}\left(  \Pi_{j1}\rho_s \Pi_{j1}\right)  ,\text{ \ }\rho
_{j2}=\frac{\Pi_{j1}\rho_s \Pi_{j1}}{p_{j}}.
\end{equation}
Thus, the quantum discord can be written as
\begin{equation}
\label{discord}
 D\left(  \rho_{s2|1}\right)=S_{\rm v-N}\left(  \rho_{s1}\right) - S_{\rm v-N}\left(  \rho_s \right)  + \inf_{\left \{  \Pi
_{j1}\right \}  }\sum_{j}p_{j}S_{\rm v-N}\left(  \rho_{j2}\right),
\end{equation}
with the element of the projective measurement $\Pi_{j1}(j=1,2)$ and density matrix  in the Bloch representation $\rho_s'$ defined as
\begin{eqnarray}
\Pi_{j1} & = &  \frac{1}{2}\left(\mathbbm{l}+\vec{n}_j\cdot\vec{\sigma}_1\right),  \nonumber \\
\rho'_s & = & \frac{1}{4}\left(\mathbbm{1}\otimes \mathbbm{1} +\vec{a}\cdot\vec{\sigma}_1\otimes \mathbbm{1}+\mathbbm{1}\otimes \vec{b}\cdot\vec{\sigma}_2 \right.\nonumber \\
&+&\left.\sum_{i,j=1}^3T_{ij}\sigma_{i1}\otimes\sigma_{j2}\right).
\end{eqnarray}
Where $\vec{n}_1=-\vec{n}_2=(\sin\theta\cos\phi, \sin\theta\sin\phi, \cos\theta)$ is a three-dimensional unit vector in an arbitrary direction, $\vec{\sigma}=(\sigma_x,\sigma_y, \sigma_z)$ denotes a vector of Pauli matrices, $\vec{a}=\text{Tr}(\rho_s\vec{\sigma}_1\otimes \mathbbm{1})$ as well as $ \vec{b}=\text{Tr}(\rho_s\mathbbm{1}\otimes\vec{\sigma}_2)$ represents a local Bloch vector, and $T_{ij}=\text{Tr}(\rho_s\sigma_{i1}\otimes\sigma_{j2})$ denotes one component of the correlation tensor. By scanning all possible measurements with parameters $(\theta, \phi)$, the quantum discord $D$ is obtained by means of the minimization procedure.

\begin{figure*}[htb]
\epsfysize=7cm \epsfclipoff \fboxsep=0pt
\setlength{\unitlength}{1.cm}
\begin{picture}(10,7)(0,0)
\put(-3.0, 0.0){{\epsffile{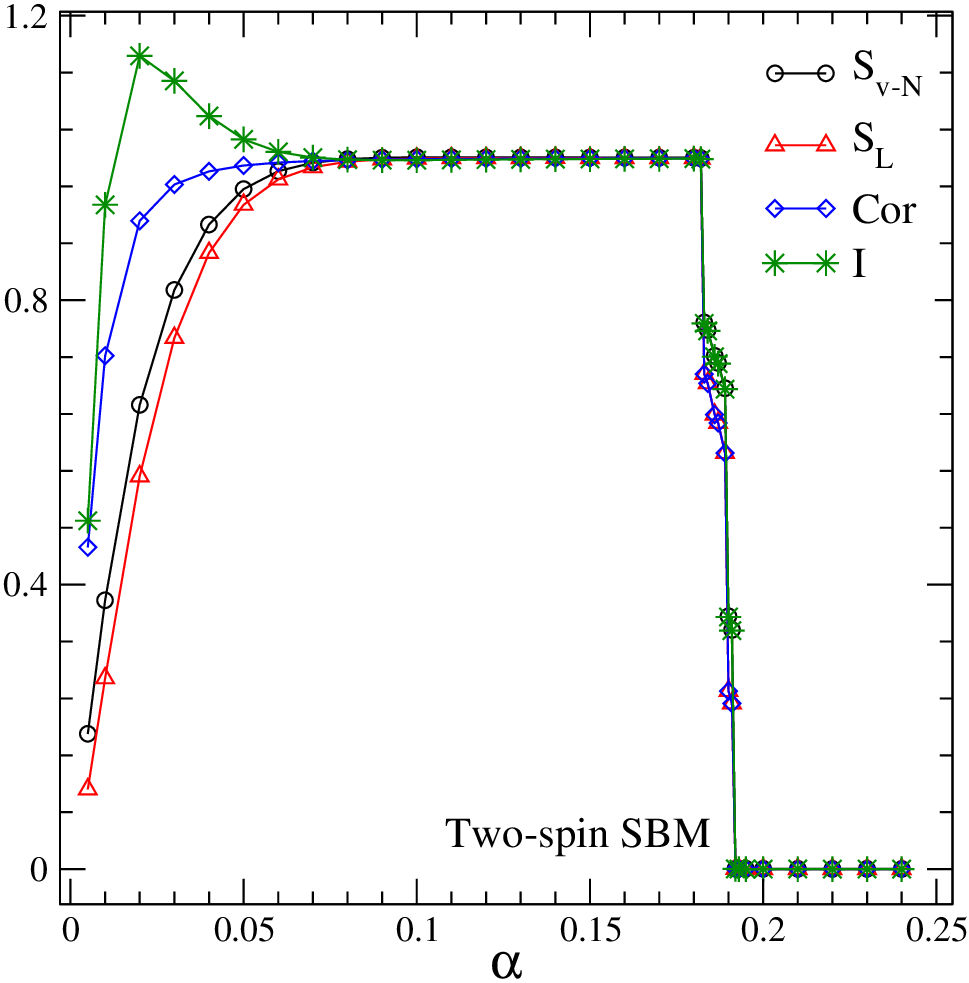}}}\epsfysize=7cm
\put(5.2, 0.0){{\epsffile{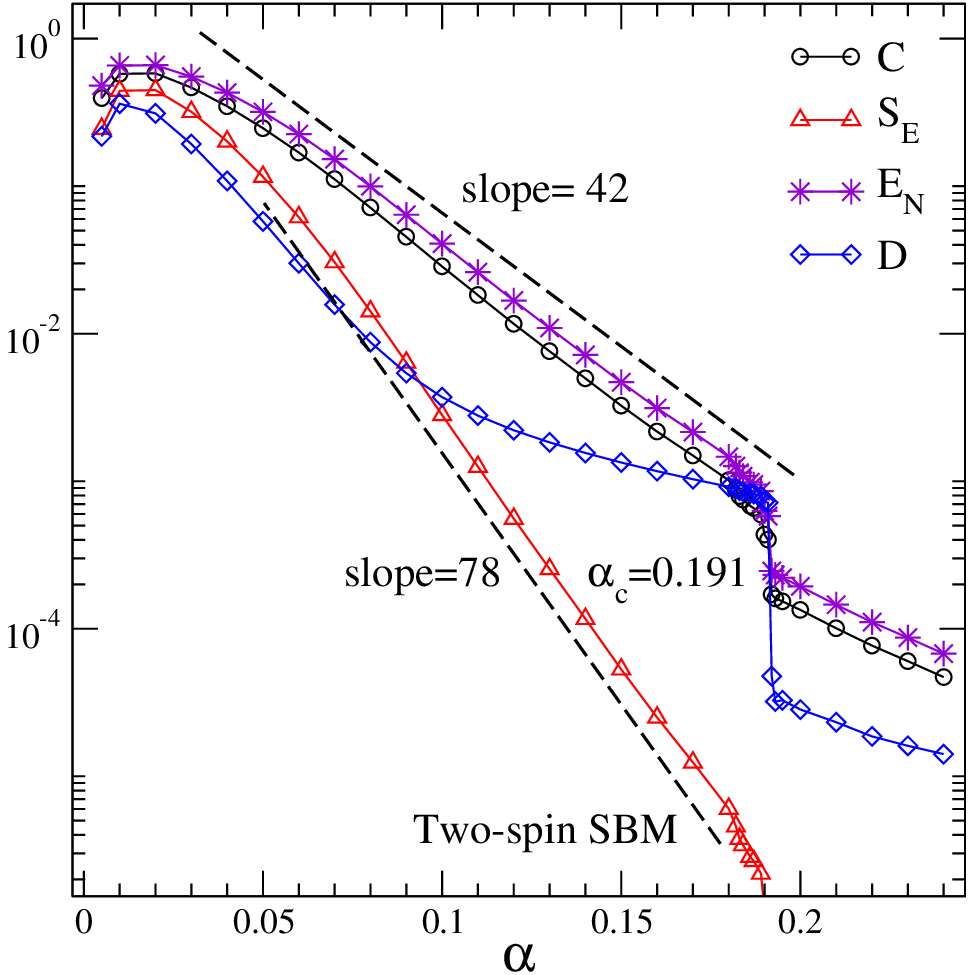}}}
\end{picture}

\hspace{0.0cm}\footnotesize{(a)}\hspace{8.0cm}\footnotesize{(b)}
\caption{ By tracing out the bath degrees of freedom, the quantum correlation, entanglement, and entropy between two spins are displayed, including the von-Neumann entropy $S_{\rm v-N}$, linear entropy $S_{\rm L}$, correlation function $\rm Cor$, and mutual information $I$ in (a) on a linear scale, and the concurrence $C$, entanglement of the formation $S_{\rm E}$, logarithmic negativity $E_{\rm N}$, and quantum discord $\rm D$ in (b) on a log-linear scale. Dashed lines represent fits to the exponential damping.
} \label{f7}
\end{figure*}

The spin-related observables defined in Eqs.~(\ref{concurrence})-(\ref{discord}) are displayed in Fig.~\ref{f7}, whose behaviors are quite different from those of bath-related ones in Fig.~\ref{f6}(a). In particular, the von Neumann entropy $S_{\rm{v-N}}$  increases monotonically due to the suppression of the renormalized tunneling amplitude, and reaches a plateau at $\alpha_t \approx 0.1$ with the maximal system-bath entanglement. It indicates that coherence is lost already before the system becomes localized, and the spin dynamics is incoherent in the plateau. This coherent-to-incoherent crossover in the two-spin model occurs at the Toulouse point $\alpha_t=\alpha_c/2$, the same as that in the single-spin model. Moreover, the entanglement between the quantum system and bath can also be measured by the linear entropy $S_L$ which behaves similarly to the von Neumann entropy $S_{\rm{v-N}}$, although the value of $S_L$ is slightly lower when $\alpha < \alpha_t$.

Besides the entropy, the spin-spin correlation function $\rm Cor$ and quantum mutual information $I$ are also plotted in Fig.~\ref{f7}(a). Since the mutual information measures the total amount of the correlations in the spin system, the relation $\rm Cor < I$ in the coupling regime $\alpha < \alpha_t$ indicates that the correlation function $\rm Cor=\langle\sigma_{z1}\sigma_{z2}\rangle - \langle\sigma_{z1}\rangle \langle \sigma_{z2}\rangle$ is an incomplete measure of the correlation. Further analysis points out that $\rm Cor$ has a similar behavior to $C\left(  \rho_{s2}:\rho_{s1}\right) = I - D$, suggesting it belongs to the degree of the classical correlation. Besides, it reaches its maximum  $\rm Cor=I=1$ at $\alpha \ge \alpha_t $, the same as the system-bath entanglement $S_{\rm{v-N}}$. Therefore, one can conjecture that the spin-spin correlation $\rm Cor$ may come from the effect of the bath-induced ferromagnetic interaction $K_r=(-4\alpha\omega_c/s)$ \cite{zho18}.

In Fig.~\ref{f7}(b), the entanglement between two spins is depicted by the concurrence $C$, entanglement of formation $S_{\rm E}$, and logarithmic negativity $E_{\rm N}$. Exponential decays of them are found in the delocalized phase with large slopes, i.e., $42$ for $E_{\rm N}$ and $C$, and $78$ for $S_{\rm E}$, respectively. It indicates that the entanglement diminishes rapidly with the environmental coupling $\alpha$. Besides, the quantum discord $D$ reflecting the nonclassical correlation is also plotted in this subfigure. In contrast to the mutual information $I$ and spin-spin correlation function $\rm Cor$, quantum discord $D$ exhibits a monotonic smooth decrease in the delocalized phase. Abrupt drops of the curves are found at the transition point $\alpha_c=0.191$, analogous to the universal jump of the superfluid density in the XY model, again supporting that the quantum phase transition belongs to the Kosterlitz-Thouless universality class.

Note that the behaviors of the correlation, entanglement, and entropy between two spins significantly differ from those between two bath modes. For example, the classical correlation in the former ${\rm Cor} \sim I - D $ increases with the coupling $\alpha$, and reaches its plateau with the maximal correlation. While the one in the latter $\sum(I_{\rm b}-D_{\rm b})$ vanishes in the delocalized phase, showing the pure quantumness which can be identified as a signature of the bosonic bath. Furthermore, the logarithmic negativity $E_{\rm N,b}$ decreases with the coupling, displaying the opposite trend of the quantum discord $D_{\rm b}$. It indicates that the nonclassical correlation in the Ohmic bath is irrelevant to the entanglement, in contrast to that in quantum spin system where the discord is strongly restricted by the entanglement as shown in Fig.~\ref{f7}(b). Further analysis suggests that this nonclassical correlation is essentially determined by the quantum correlation in the position space $\rm Cor_X$, as given in Fig.~\ref{f6}(b). Although the average entanglement in the bath $\sum E_{\rm N,b} / (M-1)$ is negligibly small as compared to that $E_{\rm N}$ between two spins, both of them decay with the coupling $\alpha$ in the delocalized phase. At the transition point, however, there exist an abnormal increase of $\sum E_{\rm N,b}$ and an abrupt drop of $E_{\rm N}$, suggesting different singularities of quantum entanglements.

\subsection{First-order phase transitions}
\begin{figure*}[htb]
\centering
\epsfysize=8cm \epsfclipoff \fboxsep=0pt
\setlength{\unitlength}{1.cm}
\begin{picture}(9,8)(0,0)
\put(0.0,0.0){{\epsffile{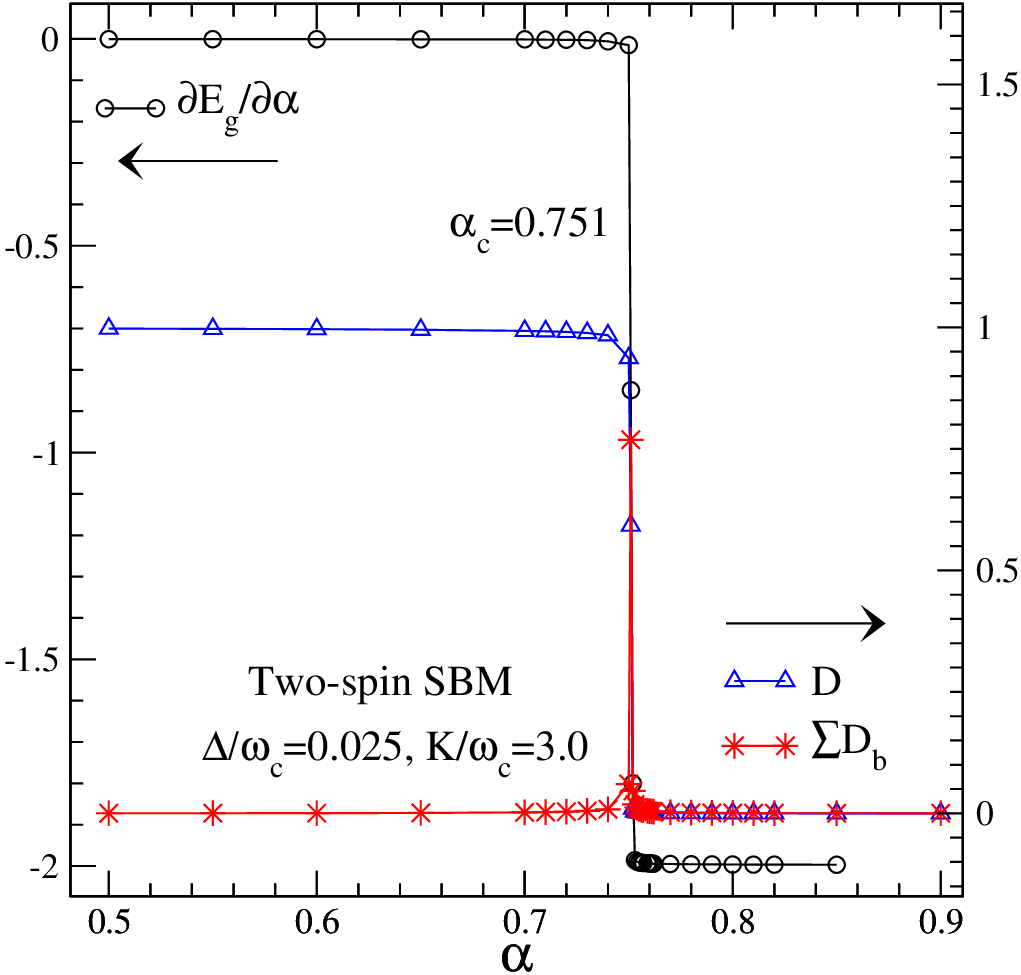}}}
\end{picture}
\caption{ In the two-spin SBM with the Ising-type interaction $K=3.0\omega_c$, the ground-state energy derivative $\partial E_g/\partial \alpha$ as well as the quantum discords $\rm D$ and $\rm D_{\rm b}$ for the quantum spin system and its bosonic environment, respectively, is displayed on a linear scale. The tunneling constant $\Delta =0.025\omega_c$, numbers of coherent-superposition states and bath modes $N=6, M=1000$, and spectral exponent $s=1$ are set. The left and right arrows indicate the Y coordinates for the energy derivative $\partial E_g/\partial \alpha$ and quantum discords $\rm D$ and $\rm D_{\rm b}$, respectively. }
\label{f8}
\end{figure*}

In this subsection, ground-state properties of the two-spin model under a strong antiferromagnetic coupling are investigated, taking $K = 3.0$ as an example. The numbers of coherent-superposition states $N$ and effective bath modes $M$ are set to be the same as those in the subsection ``B''.
In Fig.~\ref{f8}, the discontinuity in the first derivative of the ground-state energy, $\partial E_g/\partial \alpha$, is obtained at $\alpha_c=0.751(2)$, pointing to a first-order phase transition. The spin-related quantum discord $D$ and summation of the quantum discord $\rm \sum D_b$ are then plotted as representatives to illustrate quantum correlations in the spin system and bosonic bath, respectively. In the whole delocalized phase, the quantum discord $D$ remains unchanged, quite different from that in the case with $K=0$ as shown in Fig.~\ref{f7}(b). An abrupt jump from $D=1$ to $0$ is found at the transition point $\alpha_c$, yielding a singularity in the quantum discord. Further studies give the coincidence between the quantum discord and entropy of entanglement, indicating a pure entangled state of the spin system. Moreover, the relation $\langle\sigma_{z1}\sigma_{z2}\rangle = -1$ leads to an antiparallel spin configuration. Therefore, the ground state of the spin system  in the delocalized phase can be approximated as one Bell basis of maximally entangled states, $\sqrt{2}/2(|+-\rangle + |-+\rangle)$. In the localized phase, however, both the classical and nonclassical correlations vanish at $\alpha > \alpha_c$, showing the independence of two spins.

In contrast to the spin system, the nonclassical correlation $\rm \sum D_b$ of the bosonic bath vanishes in both of the localized and delocalized phases, and possesses a delta-function peak at the transition. In addition, this delta-function singularity is also found for the summations of the correlation function $\rm Cor_X$, von-Neumann entropy $\rm S_b$, mutual information $\rm I_b$, and linear entropy $S_{\rm L,b}$, thereby one concludes that arbitrary two bath modes are always independent except at the transition point. Interestingly, the system-bath entanglement represented by the von Neumann entropy $S_{\rm{v-N}}$ behaves the same as $\rm \sum D_b$, indicating that the sharp drop of the entanglement between two spins $E_{\rm N}$ is triggered by the emergence of such singularity in the Ohmic bath.

\section{Conclusions} \label{sec:con}
Quantum entanglement and correlation of bosonic baths in dissipative quantum systems have been numerically studied comprehensively based on the variational principle for ground-state phase transitions, taking the spin-boson model in a high dense spectrum $\Lambda \rightarrow 1$ and a broad range of frequency $\omega_{c}/\omega_{\rm min} $. Since phase diagrams are rich, four different cases are considered, which are the single-spin one in sub-Ohmic regime, Ohmic ones in single-spin and two-spin models, and two-spin one with a strong antiferromagnetic coupling. By comparing and analyzing several measures borrowed from quantum information theory, delta-function, cusp-like, and discontinuous singularities have been obtained, corresponding to quantum phase transitions of first-order, second-order, and Kosterlitz-Thouless types, respectively. Besides, the values of transition points and critical exponents have been accurately determined, which are much better than previous variational results, and in good agreement with analytical predictions and results from other numerical approaches.

Offering the bath-related quantum discord as a representative example, the frequency dependence has been carefully examined. Scaling form of the discord $D_{\rm b}(\omega_k, \alpha)$ has been confirmed by the data collapse technique for the Ohmic case, yielding an exponentially divergent correlation length. It is in contrast to that in the sub-Ohmic case where all curves of $D_{\rm b}(\omega_k, \alpha)$ for different couplings exhibit sharp peaks at almost the same frequency. It indicates that they belong to different universality classes. In the two-spin model, the investigation of the correlation, entanglement, and entropy between two spins have also been carried out for comparison. The behaviors of them show a great difference, compared to those between two bath modes. Specifically speaking, the classical correlation in the former increases with the coupling, and reaches its plateau at the Toulouse point, while the latter vanishes over the whole $\alpha$ range, showing pure quantumness of the bosonic bath. In addition, the quantum discord between two spins is related to the quantum entanglement, and the bath-related one is decided by the correlation function in the position space rather than the entanglement. Finally, quantum entanglements behave quite differently in the spin system and bosonic bath, though they both diminish rapidly with environmental couplings.

For discontinuous transitions, two spins are maximally entangled in the delocalized phase, and independent in the localized phase. In contrast, bath modes are independent on both two sides except at the transition point, pointing to a delta-function singularity. Further studies indicate that the sharp drop of the entanglement between two spins may be triggered by the emergence of such singularity from the bosonic bath.

{\bf Acknowledgements:} This work was supported in part by National Natural Science Foundation of China under Grant Nos. $11875120, 11775065$, and $12175052$.

\bibliography{sbm}
\bibliographystyle{apsrev4-1}

\end{document}